\documentclass[aps,pre,twocolumn,bibnotes,groupedaddress]{revtex4-1}
\usepackage{amsmath,amsthm,amsfonts,amssymb,bm}
\usepackage{graphicx,subfigure}
\usepackage[dvipsnames]{xcolor}
\usepackage[]{natbib}

\usepackage[colorlinks=true,citecolor=blue]{hyperref}

\usepackage{epsfig}
\usepackage{epstopdf}

\numberwithin{equation}{section}

\newcommand{\vecv}[1]{\bm{{#1}}}
\newcommand{\tens}[1]{\bm{{#1}}}

\newcommand{\partials}[2]{\partial_{#2}{#1}}

\newcommand{\be}{\begin{equation}}
\newcommand{\ee}{\end{equation}}
\newcommand{\bea}{\begin{eqnarray}}
\newcommand{\eea}{\end{eqnarray}}
\newcommand{\EC}{\partials{F}{\epsilon}_{\rm elastic}}

\newcommand{\tw}{t_{\rm w}}

\newcommand{\edotbar} { \bar{\dot{\epsilon}}}
\newcommand{\ebar} { \bar{\epsilon}}
\newcommand{\edot} { \dot{\epsilon}}
\newcommand{\etae} { \eta_{\rm E}}
\newcommand{\sigmae} { \sigma_{\rm E}}
\newcommand{\sigmaeb} { \sigma_{\rm E0}}
\newcommand{\force} { F}
\newcommand{\ftrans}{\tilde{F}}
\newcommand{\considere} {Consid\`ere }

\newcommand{\taur}{\tau_R}
\newcommand{\taud}{\tau_d}
\newcommand{\taus}{\tau_s}
\newcommand{\taub}{\tau_b}

\newcommand{\conf}{\tens{W}} 
\newcommand{\total}{\tens{T}}
\newcommand{\visc}{\tens{\Sigma}}
\newcommand{\Sfunc}{\tens{S}}
\newcommand{\evolve}{\tens{f}}

\newcommand{\dS} {  \dot{\sigma}_{\rm  E}}
\newcommand{\dSb} { \dot{\sigma}_{\rm  E0}}
\newcommand{\ddS} { \ddot{\sigma}_{\rm E}}
\newcommand{\ddSb} {\ddot{\sigma}_{\rm E0}}

\begin{document}
\title{Criteria for extensional necking instability in complex fluids
  and soft solids. \newline Part I: imposed Hencky strain rate protocol.}

\author{D. M. Hoyle} \email{d.m.hoyle@durham.ac.uk}
\homepage{http://community.dur.ac.uk/d.m.hoyle/}
\affiliation{Department of Physics, University of Durham, Science Laboratories, South Road,
  Durham, DH1 3LE, United Kingdom}

\author{S. M. Fielding} \email{suzanne.fielding@durham.ac.uk}
\homepage{http://community.dur.ac.uk/suzanne.fielding/}
\affiliation{Department of Physics, University of Durham, Science
  Laboratories, South Road, Durham, DH1 3LE, United Kingdom}

\begin{abstract}
  We study theoretically the necking dynamics of a filament of complex
  fluid or soft solid in uniaxial tensile stretching at constant
  imposed Hencky strain rate $\edot$, by means of linear stability
  analysis and nonlinear (slender filament) simulations. We
  demonstrate necking to be an intrinsic flow instability that arises
  as an inevitable consequence of the constitutive behaviour of
  essentially any material (with a possible rare exception, which we
  outline), however carefully controlled the experimental conditions.
  We derive criteria for the onset of necking that are reportable
  simply in terms of characteristic signatures in the shapes of the
  experimentally measured rheological response functions, and should
  therefore apply universally to all materials. As evidence of their
  generality, we show them to hold numerically in six popular
  constitutive models: the Oldroyd B, Giesekus, FENE-CR, Rolie-Poly
  and Pom-pom models of polymeric fluids, and a fluidity model
  of soft glassy materials.  Two distinct modes of necking instability
  are predicted.  The first is relatively gentle, and sets in when the
  tensile stress signal first curves downward as a function of the
  time $t$ (or accumulated strain $\epsilon=\edot t$) since the
  inception of the flow.  The second is more violent, and sets in when
  a carefully defined `elastic derivative' of the tensile force first
  slopes down as a function of $t$ (or $\edot$).  In the limit of fast
  flow $\edot\tau\to\infty$, where $\tau$ is the material's
  characteristic stress relaxation time, this second mode reduces to
  the \considere criterion for necking in solids. However we show that
  the \considere criterion fails to correctly predict the onset of
  necking in any viscoelastic regime of finite imposed $\edot\tau$,
  despite being widely discussed in the complex fluids literature.
  Finally, we elucidate in detail the way in which these modes of
  instability manifest themselves in entangled polymeric fluids
  (linear polymers, wormlike micelles and branched polymers). In
  particular we demonstrate four distinct regimes of necking behaviour
  as a function of imposed strain rate, consistent with master curves
  in the experimental literature.
\end{abstract}

\date{\today}
\maketitle

\section{Introduction}
\label{sec:intro}

Extensional flows provide a crucial benchmark for constitutive
theories of the rheology of complex fluids. Under conditions of
constant imposed Hencky strain rate, material elements separate
exponentially quickly and so subject the fluid's underlying
microstructure (polymer chains, wormlike micelles, {\it etc.}) to much
more severe reorganisation than is typically experienced in shear.  In
consequence, many nonlinear flow features manifest themselves only in
extension:
{\it e.g.}, the extensional strain hardening exhibited by many
polymeric fluids.  Extensional flows are therefore highly sensitive to
the underlying fluid microstructure,
and prove important in discriminating between competing constitutive
theories.

Key aims of any constitutive model are to predict a fluid's stress
response as a function of applied strain history (or vice versa), in
both shear and extension.  In the literature, many theoretical
calculations make the simplifying assumption that the flow field
remains spatially uniform in any given flow protocol.  In practice
this assumption may prove valid, to good approximation, in some region
of a carefully designed flow cell. Near uniform simple shear may
obtain in a narrow gap Couette device, for example. Likewise a region
of near uniform extension may obtain in the vicinity of the central
stagnation point of a cross slot device~\cite{Hoyle2013a,Haward2012},
or in the central part of a hyperbolic contraction flow
cell~\cite{Galindo-Rosales2013}.

However for many complex fluids, in many flow situations, conditions
of uniform rheometric flow prove unattainable -- or at least
unsustainable -- as a matter of fundamental principle, even in the
most carefully designed flow devices, as an unavoidable consequence of
a flow instability inherent to the fluid's constitutive response.
Under an imposed shear, for example, the phenomenon of shear banding
arises widely as a consequence of a flow instability associated with a
regime of negative slope in the underlying constitutive curve of shear
stress as a function of shear rate. When performing calculations to
compare with bulk measurements, it is then crucial to take this
heterogeneity into account by removing any assumption of uniform flow
in the calculations.  Experimentally, spatially resolved velocimetry
becomes necessary to ascertain the local constitutive response in each
band separately.

To characterise a fluid's extensional rheology, a common experiment
consists of stretching out in length an initially undeformed
cylindrical filament (or rectangular sheet) of the material.  For a
review of filament stretching techniques, see~\cite{McKinley2002b} and
references therein.  In stretching at constant imposed Hencky strain
rate $\edot$, the extensional stress growth coefficient $\etae^+(t,\edot)$ records the extensional
(tensile) stress $\sigmae^+(t,\edot)$, normalised by $\edot$, as a function of
the time $t$ (or accumulated strain $\epsilon=\edot t$) since the
inception of the flow.  If this quantity can be measured to steady
state, in some part of the sample at least, the steady tensile stress
plotted as a function of applied strain rate, obtained in a series of
stretching experiments performed at different strain rates, then gives
the extensional constitutive curve, or flow curve, $\sigmae(\edot)$.

Another common protocol comprises stretching a filament at constant
tensile stress
$\sigmae$~\cite{Munstedt1975,Munstedt2013,Munstedt2014}.  This
typically allows a fluid to attain a steady flow more quickly than
under conditions of constant strain rate, giving readier access to the
constitutive curve \cite{Alvarez2013}.  Stretching at constant
tensile force $\force$ provides a more natural mimic of some industrial
processes, such as fibre spinning \cite{Wagner2002,Szabo2012}.

Even in the most carefully performed filament stretching
experiments, however, almost ubiquitously observed is the onset of
heterogeneous deformation. Typically (in a filament stretching
rheometer at least) the central region of the filament, furthest from
the sample ends, develops a higher strain rate than the globally
applied one and thins more quickly than the sample as a whole. This
eventually causes the filament to fail altogether, aborting the
experimental run.  It has been seen in linear polymers
\cite{Barroso2005a}, branched polymers \cite{Liu2013,Burghelea2011a},
associative polymers \cite{Tripathi2006}, wormlike micelles
\cite{Bhardwaj2007}, bubble rafts \cite{Arciniaga2011}, and dense
colloidal suspensions \cite{Smith2010}. It arises in all common
stretching protocols, including at constant tensile stress
\cite{Andrade2011}, constant applied Hencky strain rate
\cite{Burghelea2011a,Malkin2014}, and following a finite Hencky strain
ramp~\cite{Wang2007}).

Two qualitatively different modes of failure are widely
reported~\cite{Barroso2005,
  Barroso2010,Wang2008,Wang2010,Wang2011a,Liu2013a}. In slow
stretching, failure typically occurs via a process of ductile necking
in which the variations in cross sectional area that develop along the
filament's length are relatively gradual.  Faster stretching
experiments typically cause more dramatic failure in which the sample
sharply rips across its cross section in a `rupture' or
  `fracture' event. (Indeed, it is unclear whether there exists a
  distinction between the `rupture' that is often refered to
  experimentally, and `fracture'. We prefer the term `fracture' in
  refering to these more dramatic events in fast stretching.)

Experimentally, the occurrence of necking (or fracture) presents a
significant technical challenge to characterising a material's
extensional rheology. This is true even before the filament finally
fails, because the flow field becomes heterogeneous and therefore
renders the measurement of a material's local, homogeneous material
response functions more challenging. In particular, the extensional
strain rate in the necking part of the sample increases above the
globally applied one in a manner that is a priori unpredictable. A
comparison of experimental data with the results of calculations that
assume a homogeneous flow therefore becomes much more difficult.

To overcome this difficulty, feedback strategies have been developed
that monitor the strain rate in the necking region and responsively
reduce the globally applied strain accordingly, to keep the strain
rate in the neck temporally constant.  Knowledge of the tensile force
divided by the local area in the neck then gives the time-dependent
local tensile stress at the given, feedback-controlled constant local
strain rate \cite{Bach2003a,Rasmussen2005a,Alvarez2013}. Such
techniques can be thought of as the counterpart in extensional
rheometry of performing spatially resolved velocimetry in shear banded
flows, with the additional feature of actively controlling the flow by
feedback.  However successful such strategies in temporarily
stabilising a constant strain rate in the neck, though, it seems
  highly unlikely that necking could be avoided altogether by any
  feedback algorithm, being (as we shall show in what follows) a flow
  instability intrinsic to the material's constitutive behaviour.

With this experimental backdrop in mind, the aim of this work is to
study theoretically the onset of necking in filament stretching of
complex fluids and soft solids under conditions of constant imposed
Hencky strain rate.  (In a separate manuscript\cite{stress}, we consider
necking at constant imposed tensile stress or tensile force.)
Starting with a time-dependent ``base state'' corresponding to a
uniform cylinder being stretched, we perform a linear stability
analysis for the dynamics of (initially) small perturbations in the
cylinder's cross sectional area, to determine the onset of necking. We
then perform nonlinear simulations to elucidate the dynamics once the
necking heterogeneity has attained a finite amplitude, beyond
the linear regime. For definiteness we consider a cylindrical filament
in uniaxial stretching, though the criteria that we shall derive also
apply to planar extension.

Our objectives are threefold. First, we seek to demonstrate the
phenomenon of necking to be a flow instability that arises as an
inevitable consequence of the constitutive behaviour of essentially
any complex fluid or soft solid, however carefully the experiment
  is performed, unavoidably leading to a heterogeneous extensional
flow field along the filament.  (In this sense, necking can be viewed
as an extensional counterpart of the banding instability that is
widely seen in shear.)

Second, we derive criteria for the onset of necking that are universal
to all complex fluids and soft solids, and are reportable simply in
terms of characteristic signatures in the shapes of the experimentally
measured material response functions of tensile stress $\sigmae^+$
and/or tensile force $F$ as a function of the time (or accumulated
strain) since the inception of the flow. We shall first derive these
by means of analytical linear stability calculations performed within
a constitutive model of a highly generalised form.  We then confirm
their generality by numerical simulations in six concrete choices of
constitutive model: the phenomenological Oldroyd B, Giesekus and
FENE-CR models; the microscopically motivated Rolie-Poly model of
entangled linear polymers and wormlike micelles; the Pom-pom model of
entangled branched polymers; and a simplified fluidity model of soft
glassy materials (foams, emulsions, dense colloids, microgels, {\it
  etc.}), which display a yield stress and rheological ageing
\cite{Fielding1999,Sollich1996,Cates2003}.

Third, we seek to elucidate the way in which our universal criteria
manifest themselves in those three major classes of complex fluids: in
entangled linear polymers and wormlike micelles; in entangled branched
polymers; and in soft glassy materials.  In the context of entangled
polymers, a particular objective is to demonstrate the tube theory of
polymer rheology, with chain stretch and convective constraint
release, to be capable of capturing all the features of a widely
discussed experimental master-curve of the strain at which a sample
fails as a function of imposed strain rate.  In ageing soft glassy
materials, we demonstrate that the sample fails by one of two
qualitatively different modes of necking, according to the sample age
at the time stretching commences.

Commonly discussed in the literature as a predictor for the onset of
necking is the \considere criterion
\cite{Petrie2009,Considere1885}.  This predicts necking to set in
when the tensile force attains a maximum as a function of the
accumulated strain, then subsequently declines.  It was originally put
forward in the context of solid mechanics, for which it is indeed
appropriate to take the accumulated strain $\epsilon$ as the only
relevant deformation variable. For a complex fluid, however, also
crucial is the strain {\em rate} $\edot$ at which stretching is
performed, relative to the inverse characteristic relaxation time
$1/\tau$ of the fluid in question.  In view of this there is no
reason, a priori, for the \considere criterion to apply to complex
fluids.  Indeed, our calculations will demonstrate that it performs
poorly in predicting the onset of necking at low to moderate imposed
strain rates (compared to $1/\tau$).  It does, however, compare quite
well with experiments \cite{Barroso2005,Barroso2005a} and
simulations \cite{Hassager1998,McKinley1999a} at high strain rates
$\edot$.  Indeed this is to be expected, because viscoelastic fluids
tend towards a solid-like response in the regime $\edot\tau\gg 1$.  In
important contrast, the criteria for the onset of necking offered in
this work are valid not only in this fast flow regime, but apply
across the full range of flow rates from slow to fast. We shall also
demonstrate the way in which one of our criteria reduces to the
\considere criterion in the limit of fast flow $\edot\tau\to\infty$.

An early insightful attempt to consider the importance of both strain
and strain rate in the onset of necking can be found in
\cite{Hutchinson1977}.  Stability analyses were later performed for
Newtonian and Maxwell fluids in\cite{Ide1977}, and
in the Oldroyd B and FENE-CR models in
\cite{Olagunju1999,Olagunju2011}.  A scaling theory based on a
critical recoverable strain was put forward in
\cite{Joshi2003,Joshi2004}. An empirical criterion for rupture was
offered in \cite{Malkin1997a}. Direct numerical simulations of
necking have been performed within constitutive models of polymers
\cite{Rasmussen2013,McKinley2002b,Bhat2010}, wormlike micelles
\cite{Cromer2009}, and amorphous elastoplastic solids
\cite{Eastgate2003}.

The criteria that we discuss in what follows were first outlined in
\cite{Fielding2011} in the context of polymeric fluids and in
\cite{Hoyle2015} in the context of soft glassy materials
respectively. The purpose of the present manuscript is to give a much
more detailed explanation of the criteria announced in those earlier
Letters, and comprehensive numerical evidence supporting them.

Our focus will be on the onset of necking in a highly viscoelastic
filament of sufficiently large radius that bulk viscoelastic stresses
dominate surface effects. Accordingly, we set the surface tension to
zero in most of our calculations. We therefore do not address
capillary breakup as studied in CaBeR rheometers \cite{Clasen2006,Tembely2012,Szabo2012,Vadillo2012,Webster2008,McIlroy2014,Bhat2008}. However,
towards the end of the paper we shall return to incorporate surface
tension and show that it affects our results only in the regime of very
slow strain rates.

All our calculations are performed within a slender filament
approximation in which the wavelengths of any variations along the
filament's length are taken as long compared to the filament radius.
Our approach therefore cannot capture the details of the final
pinchoff of any neck~\cite{Renardy2002,Eggers2008}, nor can it capture
a fracture mode in which the filament sharply rips across its cross
section~\cite{Ligoure2013,Rycroft2012,Eastgate2003}.  Such phenomena
are deferred to future study. It is worth noting, however, that
  we might expect the violent necking predicted below in fast flows to
  be replaced by a sharp fracture event in any fully 3D simulation
  capable of capturing this.

So far, we have discussed the onset of a heterogeneous profile along
the filament in terms of a true material instability (necking).  It is
important to note, however, that even before a true necking
instability arises, some narrowing of the central region of the
filament relative to that near each endplate is to be expected because
of the boundary condition that prevents the fluid from slipping at the
plates, and therefore prevents those parts of the sample nearest the
plates from being properly stretched.  Indeed, below we shall discuss
the way in which this initial heterogeneity induced by the flow
geometry acts as a seed that is then picked up and hugely amplitude by
the true material necking instability.  However we also note
  that, even in a thought experiment in which the effect of the
  boundary conditions could be removed altogether (for example by
  perfectly cothinning the endplates concurrently to match the
  changing diameter of the filament), the necking instability would be
  seeded by other sources of heteroeneity (such as initial sample
  imperfection) and could not be avoided. Our calculations with
  periodic boundary conditions below will confirm this.

The paper is structured as follows. We start in
Sec.~\ref{sec:preamble} with a preamble concerning measures of
extensional force and stress that are commonly reported in the
experimental literature. In Sec.~\ref{sec:models} we discuss the
constitutive models and flow protocol to be studied throughout the
paper. In Sec.~\ref{sec:LSA} we outline in general terms the procedure
of our linear stability analysis for the onset of necking.  In
Sec.~\ref{sec:criteria} we use this analysis, within a constitutive
model of highly general form, to derive fluid-universal criteria for
the onset of two qualitatively different modes of necking instability.
We discuss the rheological signature of these modes in the form of
characteristic features in the shapes of the material response
functions plotted versus the time (or accumulated strain) since the
inception of the flow.  To confirm the validity of these general
criteria, we then perform in Sec.~\ref{sec:results_lsa} numerical
calculations of the linearised necking dynamics within six widely used
constitutive models and show the criteria to indeed hold within them.
We further discuss in more detail the way the criteria manifest
themselves in three important classes of complex fluid: entangled
linear polymers and wormlike micelles, entangled branched polymers,
and soft glasses.  In Sec.~\ref{sec:NL} we perform nonlinear
simulations to study the dynamics once the neck has developed to
attain a finite amplitude.  Sec.~\ref{sec:conclusions} contains our
conclusions.

\section{Measures of force, stress and strain}
\label{sec:preamble}

Key variables measured as a function of the time $t$ (or accumulated
strain $\epsilon=\edot t$) during filament stretching at constant
imposed Hencky strain rate $\edot$ are the tensile force $F(t)$, and
the tensile stress, {\it i.e.}, the tensile force per unit cross
sectional area. In fact three different tensile stress measures are
commonly reported experimentally, according to which area variable is
used in the denominator:

(a) The engineering stress is defined at any time $t$ as the force
$F(t)$ normalised by the cross sectional area, $A$, of the
filament at the {\em start} of the run, $\sigma_{E,\rm {eng}}^+(t) =
F(t)/A(0)$.  Further dividing this by the constant value of the
imposed Hencky strain rate gives the engineering stress growth
coefficient $\eta_{E,\rm{eng}}^+(t) = F(t)/A(0)\edot$.  It is
important to note, however, that this quantity does not properly
characterise the tensile stress, because it does not allow for the
filament's (on average) exponentially decreasing cross sectional area
$A(t)=A(0)\exp(-\epsilon)$ as a function of the accumulating strain
$\epsilon$. In fact, the engineering stress should properly be
recognised as a measure of the time-dependence of the tensile force,
normalised by the initial area $A(0)$.

(b) The apparent tensile stress is defined at any time $t$ as the
force $F(t)$ normalised by the cross sectional area $A_{\rm
  hom}(t)=A(0) L(0)/L(t)=A(0)\exp(-\epsilon)$, as calculated by
supposing that the filament has remained perfectly uniform up to that
time, without any necking yet having occurred.  This gives an apparent
stress $\sigma_{\rm E,app}^+(t) = F(t)/A_{\rm hom}(t)$ and a
corresponding apparent stress growth coefficient $\eta_{\rm
  E,app}^+(t) = F(t)/\edot A_{\rm hom}(t)$.  As a measure of the
tensile stress this is a significant improvement on the engineering
stress, in accounting for the overall exponential decrease in the
filament's cross sectional area. However it still does not properly
report the tensile stress once necking occurs, because the cross
sectional area then varies along the filament's length.

(c) The true tensile stress at any time $t$ and location $z$ along the
filament's length is defined as the force $F(t)$ divided by the actual
cross sectional area of the filament at that point, $A(z,t)$, giving
$\sigma_{\rm E}^+(z,t)=F(t)/A(z,t)$.  (Note that although the {\em
  stress} may vary as a function of position along the filament's
length, the {\em force} must remain uniform by force balance at the
low Reynolds number flows of interest here.) The corresponding true
stress growth coefficient $\eta_{\rm E}^+(z,t)=F(t)/\edot A(z,t)$. To
track the evolution of this quantity at some location $z$ requires not
only a measurement of the tensile force, but also of the evolving
cross sectional area at that location
\cite{Rasmussen2005,Bach2003a,Alvarez2013}.

In seeking to compare our numerical results with experimental data, we
shall sometimes show data for the apparent stress growth coefficient
$\eta^+_{\rm E,app}(t)$, to make contact with experiments that do not
explicitly track the effect of necking on the cross sectional area;
and sometimes for the true stress growth coefficient $\eta_{\rm
  E}^+(z_{\rm mid},t)$, to make contact with experiments that do track
the time-evolution of the cross sectional area in the developing neck.
We use the value $z_{\rm mid}=L(t)/2$ because in all our simulations
the neck develops at the filament's midpoint.

Finally, in any experiment where necking occurs the strain and
  strain rate will vary along the filament: $\epsilon=\epsilon(z,t)$
  and $\edot=\edot(z,t)$. The averages of these along the filament
  correspond to the globally imposed strain and strain rate, $\ebar$
  and $\edotbar$ respectively, and are often called the nominal Hencky
  strain and nominal Hencky strain rate.

\section{Models and flow geometry}
\label{sec:models}

\subsection{Mass balance and force balance}
\label{sec:balance}

We write the total stress $\total(\tens{r},t)$ at time $t$ in a fluid
element at position $\tens{r}$ as the sum of a viscoelastic
contribution $\visc(\tens{r},t)$ from the internal fluid
microstructure (polymer chains, wormlike micelles, emulsion droplets,
{\it etc.}), a Newtonian contribution of viscosity $\eta$, and an
isotropic contribution with a pressure $p(\tens{r},t)$:
\be
\total = \visc + 2 \eta \tens{D} - p\tens{I}.
\label{eqn:total_stress_tensor}
\ee
The Newtonian contribution may arise from the presence of a solvent,
and/or from any polymeric (or other viscoelastic) degrees of freedom
considered fast enough not to be ascribed their own dynamics.  The
symmetric strain rate tensor $\tens{D} = \frac{1}{2}(\tens{K} +
\tens{K}^T)$ where $K_{\alpha\beta} =
\partial_{\beta}v_{\alpha}$ and $\tens{v}(\tens{r},t)$ is the fluid
velocity field.  

We consider the creeping flow limit of zero Reynolds number, in which
the condition of force balance requires the stress field
$\total(\tens{r},t)$ to be divergence free:
\be
\vecv{\nabla}\cdot\,\total = 0.
\label{eqn:force_balance}
\ee
The pressure field $p(\tens{r},t)$ is determined by the condition that
the flow remains incompressible:
\be
\label{eqn:incomp}
\vecv{\nabla}\cdot\vecv{v} = 0.
\ee

\subsection{Constitutive models}
\label{sec:constitutive}

The viscoelastic stress $\visc$ is specified by a constitutive model
for the fluid in question. In this work we consider six widely used
constitutive models. These are set out in Appendix~\ref{app:models},
along with values of any model parameters used in our numerical
studies.  While inevitably differing in their detailed form, all of
them have the same general structure, which we now outline.

The viscoelastic stress 
\be
\visc=G\Sfunc(\conf,\lambda,Q,\cdots)
\ee
is the product of a constant modulus $G$ and a dimensionless tensorial
function $\Sfunc$ of a microstructural conformation tensor $\conf$,
together with any other microscopic variables relevant to the fluid
under consideration.  (We list these here simply as
$\lambda,Q,\cdots$.  For further details, see
Appendix~\ref{app:models}.)  For a polymeric (or wormlike micellar)
fluid, the conformation tensor $\conf$ could encode the ensemble
average dyad of the end-to-end vector of a chain or subchain,
depending on the level of description. For an emulsion or foam, it
could encode the ensemble average dyad of the interfacial normals.
The dynamics of the conformation tensor is then specified by a
differential equation of the general form
\be
\label{eqn:vece}
\partial_t\conf+\vecv{v}.\nabla\conf = \evolve(\nabla
\vecv{v},\conf,\lambda,Q\cdots),
\ee
with counterpart scalar equations for the dynamics of
$\lambda,Q,\cdots$, of the same differential form. 

The governing equations of the six constitutive models studied in this
work, all of which conform to this general structure, are set out in
Appendix~\ref{app:models}. Among the these models, the Oldroyd
B~\cite{Larson1988} and FENE-CR~\cite{Chilcott1988} models provide
phenomenological descriptions for the dynamics of the conformation
tensor $\conf$ in dilute polymer solutions (with no additional
variables $\lambda,Q,\cdots$). The Giesekus model~\cite{Larson1988} is
a generalisation of Oldroyd B, aimed at modelling more concentrated
polymeric fluids.

For a microscopically motivated description of more concentrated
solutions or melts of entangled linear polymers, we use the Rolie-Poly
model\cite{Likhtman2003}.  This also recovers the reptation-reaction
model~\cite{Cates1990} of wormlike micelles for a particular choice of
model parameters.  It is based on the tube theory of Doi and Edwards
\cite{Doi1986}, whereby a polymer chain (or wormlike micelle) is
dynamically restricted by a confining tube of topological
entanglements with the surrounding chains.  The chain then refreshes
its configuration by a process of 1D curvilinear diffusion along the
tube contour, known as reptation.  Later added to this basic
description were the additional dynamical processes of chain stretch
relaxation and convective constraint release
\cite{Marrucci1996,Ianniruberto2014,Ianniruberto2014a}
(CCR), in which the relaxation of the stretch of a test chain has the
effect of also relaxing entanglement points, thereby facilitating the
relaxation of tube orientation. The Rolie-Poly
model~\cite{Likhtman2003} incorporates these three dynamical
process into a differential constitutive equation for the dynamics of
$\conf$, as set out in Appendix~\ref{app:models}.

For a microscopically motivated description of entangled long-chain
branched polymers we use the Pom-pom model \cite{McLeish1998,Blackwell2000}. Here
the presence of polymeric arms branching off each end of a polymer
molecule's main backbone inhibits the reptation of that backbone and
promotes its stretching between the branch-points. The Pom-pom model
specifies dynamics of the conformation $\conf$ of the backbone, and of
the degree of backbone stretch, $\lambda$.

Finally we consider a phenomenological fluidity model of a broad class
of disordered soft `glassy' materials (foams, dense emulsions,
colloids, microgels, {\it etc.})
\cite{Fielding1999,Sollich1996,Cates2003}.  Common to all these are
the features of structural disorder (in a dense packing of emulsion
droplets, for example) and metastability (with the large energy
barriers involved in stretching soap films inhibiting rearrangements
of the droplets). These glassy features give rise to rheological
ageing, in which a sample becomes progressively more solid-like as a
function of the time elapsed since it was prepared.  The sustained
application of flow however halts ageing and rejuvenates the sample to
a steady state with an effective age set by the inverse flow rate. The
flow curve displays a yield stress in the limit of slow flow. As
outlined in Appendix~\ref{app:models}, our fluidity model specifies
the dynamics of a tensor $\conf$ characterising the conformation of
the droplet interfaces, and of the total interfacial area $Q$. Also
specified is an evolution equation for the ageing stress relaxation
time $\tau$. We have also checked that our results for the fluidity
model, presented below, also hold within the more sophisticated soft
glassy rheology model~\cite{Sollich1996, Cates2003, Hoyle2015}.

\subsection{Units and parameter values.}
\label{sec:units}

Throughout we adopt units of length in which the initial length of the
filament $L(0)=1$, and units of stress in which the viscoelastic
modulus $G=1$.  We use units of time in which the intrinsic relaxation
timescale of any constitutive model is equal to unity. Accordingly,
for the Oldroyd B, Giesekus and FENE-CR models we set $\tau=1$. For
the Rolie-Poly model we set $\taud=1$. For the Pom-pom model we set
$\taub=1$. For the fluidity model of soft glasses we set the
microscopic time $\tau_0=1$.  (This model's actual stress relaxation
time $\tau$ becomes highly separated from $\tau_0$ during ageing.) The
definition of these timescales can be found in
Appendix~\ref{app:models}. Values for the other model parameters, in
these units, are listed in table \ref{tab::model_params} of
Appendix~\ref{app:models}.

\subsection{Initial conditions, flow geometry and protocol.}
\label{sec:geometry}

We consider a sample of material that at some initial time $t=0$ is in
the shape of an undeformed uniform cylindrical filament of length
$L(0)$ and cross sectional area $A(0)$, with an isotropic conformation
tensor $\conf(0)=\tens{I}$.  Initial conditions for any additional
variables $\lambda,Q,\cdots$ are prescribed in
Appendix~\ref{app:models}. For all times $t>0$ the filament is then
subject to a constant applied Hencky strain rate $\edotbar$, such that
its length increases as $L(t)=L(0)\exp(\edotbar t)$.  The overbar
signifies that $\edotbar$ is the strain rate experienced by the sample
as a whole, globally averaged along its full length.  Once necking
arises, the deformation rate will locally vary along the filament's
length $z$ such that the Hencky strain rate $\edot=\edot(z,t)$, with
the average $\edotbar$ of this function along $z$ remaining constant in time.

\subsection{Slender filament approximation}
\label{sec:slender}

We adopt a slender filament approximation \cite{Forest1990,
  Olagunju1999, Denn1975}, in which the wavelengths of
any variations that develop in cross sectional area along the
filament's length are assumed large compared to the filament's radius,
and the flow variables are averaged across the filament's cross
section at any location $z$ along it.  Relevant dynamical variables
are then the cross sectional area $A(z,t)$, the area averaged fluid
velocity in the $z$ direction $V(z,t)$, and the extension rate
$\edot(z,t)=\partial_z V$. 
For clarity we drop the "$+$" superscript, usually used in Journal of 
Rheology to denote time-dependence in the extensional stress, from $\sigmae^+$ in the equations that follow.

The mass balance equation (\ref{eqn:incomp}) is then written
\be
\partials{A}{t} + V\partials{A}{z}  = -\dot\varepsilon A, \label{eqn:1Dmass} 
\ee
and the force balance condition (\ref{eqn:force_balance})  
\be
0 = \partials{F}{z},
 \label{eqn:1Dmom} 
\ee
in which the tensile force
\be
\label{eqn:tforce}
F(t)=A(z,t)\sigmae(z,t),
\ee
and the total tensile stress 
\be
\label{eqn:tstress}
\sigmae = G\left(S_{zz} - S_{xx} \right) + 3\eta\dot\varepsilon.
\ee
As before, $\Sfunc=\Sfunc(\conf,\lambda,Q,\cdots)$.  The evolution
equation for the confirmation tensor is now written as
\be
\label{eqn:confSlender}
\partial_t \conf(z,t)+V\partial_z\conf=\evolve(\edot,\conf,\lambda,Q,\cdots),
\ee
with counterpart scalar differential equations for the dynamics of
$\lambda,Q,\cdots$.

\subsection{Transformation to co-extending frame}

In the constant Hencky strain rate experiment of interest here, the
length of the filament increases exponentially in time as
$L(t)=L(0)\exp(\edotbar t)$, and the area decreases (overall) as
$A(t)=A(0)\exp(-\edotbar t)$ (subject to local variations due to
necking). This makes it convenient to make a transformation to the
coextending, cothinning frame.  Accordingly we define new variables of
length $u$, velocity $v$ and area $a$:
\begin{eqnarray}
		u &=& z\exp(-\edotbar t),	\nonumber \\
		v(u,t) &=& V(z,t)\exp(-\edotbar t),	\nonumber \\
		a(u,t) &=& A(z,t)\exp( \edotbar t).
\end{eqnarray}
The differential operators then transform as
\begin{eqnarray}
\partial_z &\longrightarrow &  \exp(-\edotbar t)\partial_u,	\\
\partial_t &\longrightarrow &  \partial_t - \edotbar u \partial_u,
\end{eqnarray}
giving the transformed equations of mass balance  
\be
\label{eqn:massT}
\partials{a}{t} + (v-\edotbar u)\partials{a}{u} = -(\edot-\edotbar)a,
\ee
and force balance 
\be
\label{eqn:forceT}
 0 = \partials{\ftrans}{u},
\ee
where the transformed tensile force 
\be
\label{eqn:stressT}
\ftrans(t)=F(t)\exp(\edotbar t)=a(u,t)\sigmae(u,t).
\ee
The tensile stress $\sigmae^+$ is given as in~\ref{eqn:tstress} above,
with $\Sfunc=\Sfunc(\conf,\lambda,Q,\cdots)$ as before. The
transformed evolution equation for the conformation tensor is written
as
\be
\label{eqn:confT}
\partials{\conf}{t}  + (v-\edotbar u)\partials{\conf}{u} = \evolve(\edot,\conf,\lambda,Q,\cdots),
\ee
with counterpart scalar differential equations for any additional
variables $\lambda, Q,\cdots$.

\subsection{Simplified scalar model}
\label{sec:toy}

So far, we have outlined the full tensorial constitutive models to be
used in our numerical calculations, within a slender filament
approximation. We shall also perform analytical calculations within a
simplified scalar model that considers only the (assumed) dominant
component $Z=W_{zz}$ of microstructural deformation that develops in a
filament stretching experiment, also at the level of slender filament.
Conditions of mass balance and force balance remain as in
Eqns.~\ref{eqn:massT} to~\ref{eqn:stressT} above.  We then write the
tensile stress simply as
\be
\label{eqn:stressToy}
\sigmae=G Z + \eta\edot,
\ee
with the dynamics of $Z$ specified as
\be
\partials{Z}{t}+V\partials{Z}{z}=\edot f(Z)-\frac{1}{\tau}g(Z),
\ee
with separate loading and relaxation dynamics characterised by the
functions $f$ and $g$ respectively. For notational simplicity, in
this scalar model, we have also absorbed a factor $3$ into the solvent
viscosity $\eta$.

Writing the model in this highly generalised form, without specifying
any particular functional forms for the loading and relaxation
dynamics $f(Z)$ and $g(Z)$, will enable us to derive criteria for the
onset of necking that are reportable simply in terms of characteristic
signatures in the shapes of the material response functions (tensile
stress, {\it etc.}) as a function of the time (or accumulated strain)
since the inception of the flow.

\subsection{Boundary conditions}

In our linear stability calculations we assume periodic boundary
conditions between the two ends of the filament, thereby implicitly
taking the filament to correspond to a torus being stretched.  We
performed our nonlinear numerical simulations for two different sets
of boundary conditions in turn: first, we adopted the periodic
boundary conditions just discussed; and second, we used an approximate
mimic of the no-slip boundary condition between the fluid and the
endplates. (The full flow field near the plates cannot however be
captured at the level of this slender filament calculation.)

The second of these conditions is discussed in detail in
Sec.~\ref{sec:NL}. As we shall demonstrate, it automatically provides
some heterogeneity that seeds the formation of a neck in the sample.
For the nonlinear simulations performed with the periodic boundary we
instead seeded the instability by adding a small initial perturbation
to the area profile such that $a(u,t=0) = a(0) + \delta a_0 \cos(2\pi
u)$ with $\delta a_0 \ll 1$. In fact we found the necking dynamics
predicted with these two boundary conditions to be essentially the
same, so shall present the results of our nonlinear simulations only
with the second condition, mimicking no-slip.

\section{Linear stability analysis: general procedure}
\label{sec:LSA}

We now outline in general terms the procedure of performing a linear
stability analysis for the onset of necking.  We start by considering
a homogeneous ``base state'' corresponding to a filament that remains
in a uniform cylindrical shape as it is stretched out, with the flow
variables homogeneous along it.  To this base state are then added
small amplitude perturbations that are heterogeneous along the
filament's length, corresponding to the precursor of a neck.
Expanding the governing equations to first order in the amplitude of
these perturbations gives linearised equations for the dynamics of the
perturbations. Integrating these in time, and/or examining the
eigenvalues of the matrix that governs the linear equations, then
determines whether, and at what time during the filament stretching
process the perturbations grow and thereby take the system towards a
necked state, or whether they decay to leave a uniform filament.

\subsection{Homogeneous base state}

We consider first a uniform ``base state'', labelled with a subscript
$0$, corresponding to a filament that remains a uniform cylinder as it
is stretched out, with all flow variables homogeneous along it. In
this state the strain rate $\edot(u,t) = \edot_0= \edotbar$, the
transformed velocity $v(u,t) =v_0= u\bar{\dot\varepsilon}$ and the
transformed area $a(u,t) = a_0=A(0)$.  (In the laboratory frame, the
area thins as $A(t)=A(0)\exp(-\edotbar t)$.)  The viscoelastic
variables, for which no frame transformation is needed, follow as
homogeneous solutions $\conf_0(t),\lambda_0(t),Q_0(t),\cdots$ of their
respective equations of motion (\ref{eqn:confSlender} and its
counterparts for $\lambda,Q,\cdots$) for times $t>0$, subject to the
initial condition of the filament having been undeformed priori to
time $t=0$.  The time-dependent tensile stress
\be
\label{eqn:tstressb}
\sigmaeb(\edot,t) = G\left(S_{zz0} - S_{xx0} \right) + 3\eta\edotbar,
\ee
with 
\be
\Sfunc_0=\Sfunc(\conf_0(t),\lambda_0(t),Q_0(t),\cdots).
\ee

\begin{figure*}
	\centering
	\includegraphics[height = .90\textheight]{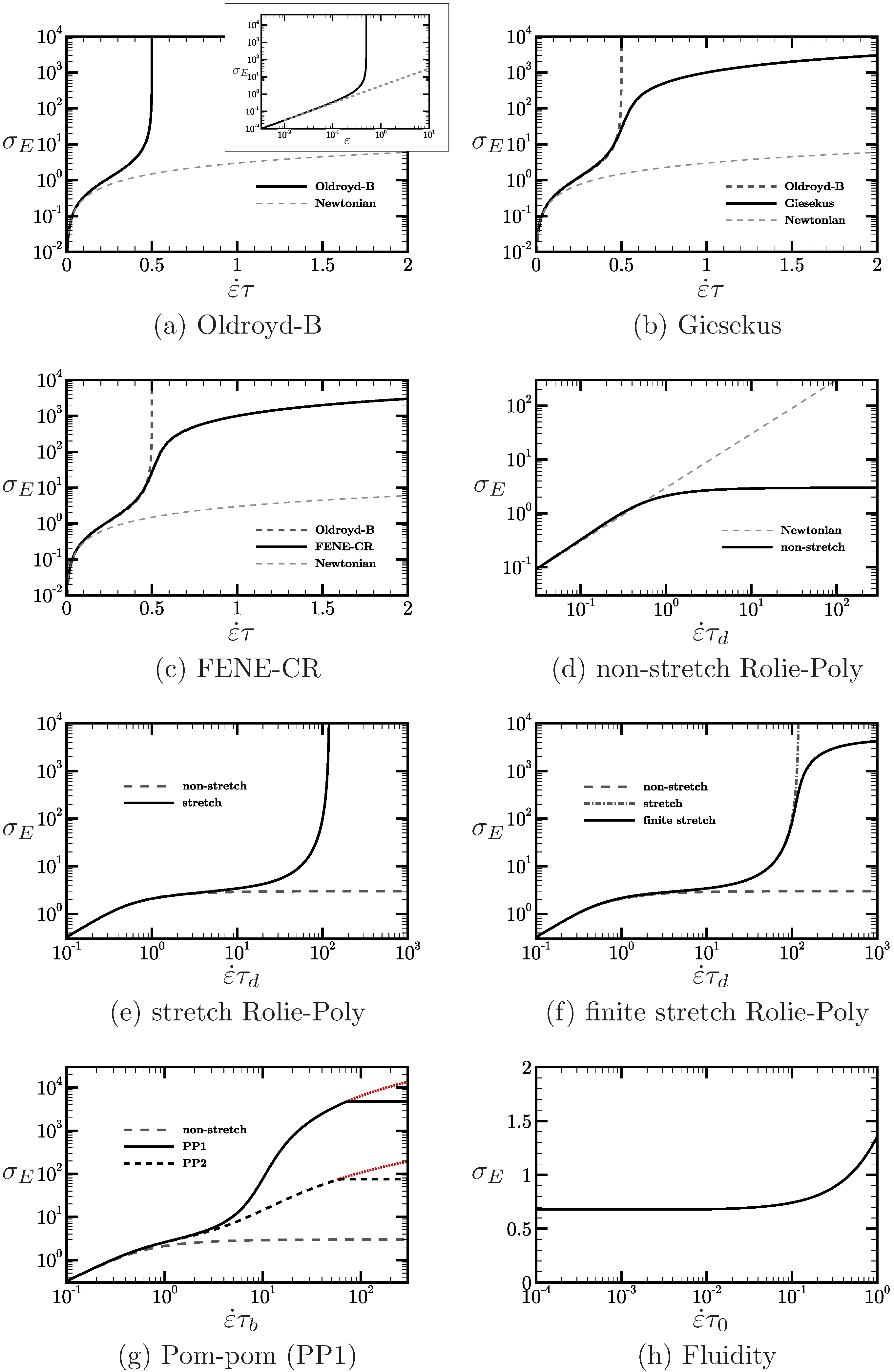}	
	\caption{Stationary homogeneous extensional constitutive
		curves of the models to be studied, for the model parameters
		values specified in App.~\ref{app:models}, table
		\ref{tab::model_params}.  The apparent negative curvature
		at low strain rates in (a), (b) and (c) is a consequence of
		the log-linear scale. The more familiar log-log scale is shown
		as an inset for the Oldroyd B model in (a). In (g) the black
		curves show the constitutive curves for the form of the
		Pom-pom model in which the backbone stretch has a hard
		cutoff. The red-dotted lines show the equivalent curves for
		a form of the model in which that cutoff is removed. See the discussion at the end of Sec.~\ref{sec:branched}.
		\label{fig:constitutive}}
\end{figure*}

If the tensile stress attains a steady state in the limit of long
times $t\to\infty$ after the inception of the flow, once many strain
units $\ebar=\edotbar t$ have been applied, the steady state relation
$\sigmaeb(\edotbar)$ defines the material's homogeneous extensional
constitutive curve. The constitutive curves of the six models
considered in this work are shown in Fig.~\ref{fig:constitutive}. We
shall return in the results sections below to describe the shapes of
these curves in more detail, in particular discussing any features
that pertain to necking.

\begin{figure*}
	\centering
	\includegraphics[]{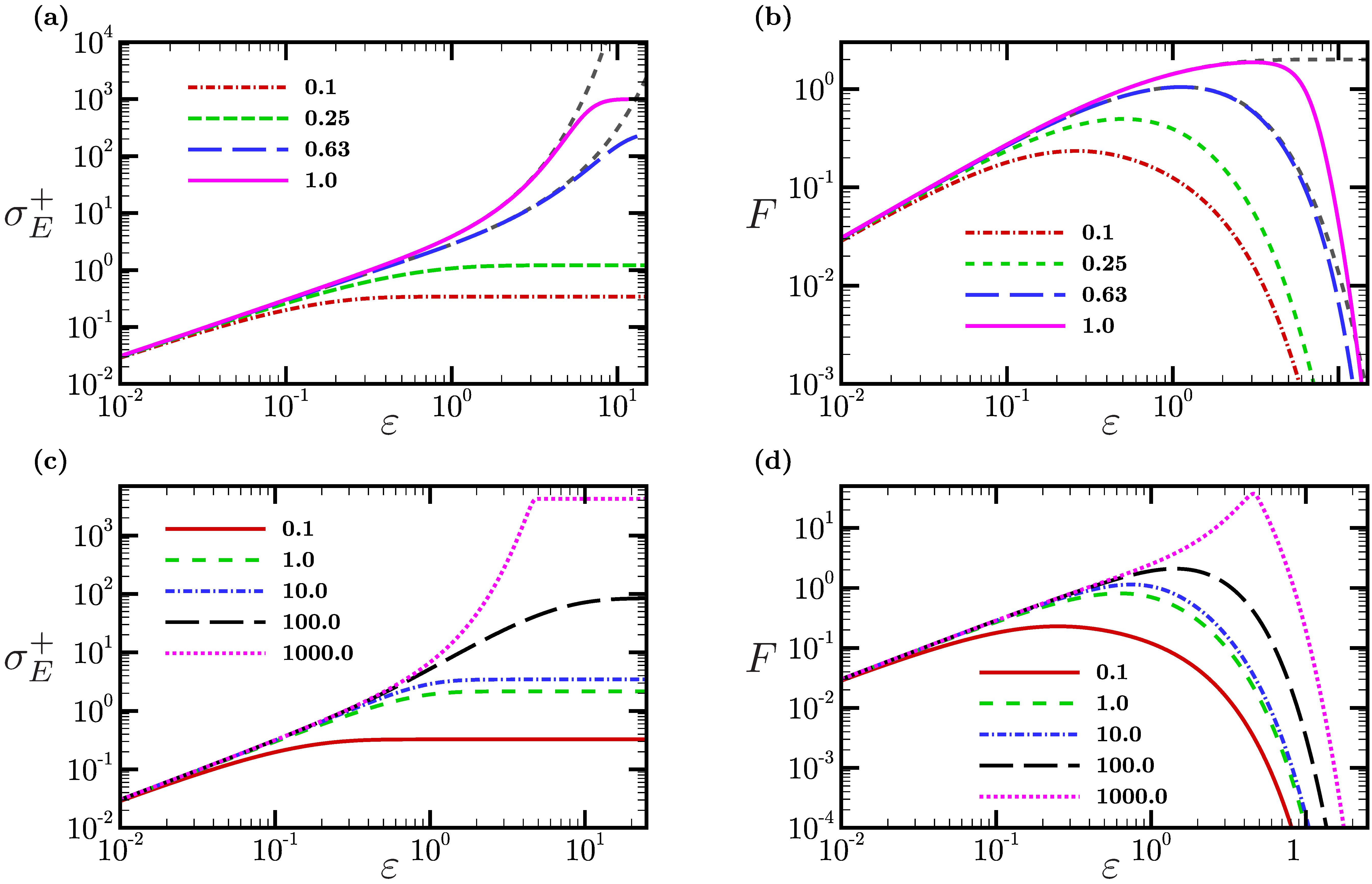}	
	\caption{Transient evolution of the stress $\sigmae^+$ (left) and  the force $F$ (right) for the Giesekus model (top) and the finite-stretch Rolie-Poly model (bottom), within a calculation that artificially constrains the flow to remain homogeneous. The key shows the strain rate for each data set. The counterpart steady-state homogeneous constitutive curves $\sigmae(\edot,t\to\infty)$ can be seen in figure \ref{fig:constitutive}. Parameter values for each model are given in appendix \ref{app:models}, table \ref{tab::model_params}. The grey dashed curves in the Giesekus (top) figures show for comparison results for the Oldroyd B model, to which the Giesekus model reduces in the limit $\alpha\to 0$.
          \label{fig:stress}}
\end{figure*}

The time-evolution $\sigmaeb^+(t)$ of the homogeneous stress signal
towards this steady state constitutive curve is shown at several
different imposed strain rates for the Giesekus and Rolie-Poly models
in the left panels of Fig.~\ref{fig:stress}. The corresponding
(untransformed) tensile force $F(t)=\sigmae^+(t)A(t)$ is shown in the
right panels of the same figure. The force initially increases, due to
the rising stress, then later decreases, due to the declining cross
sectional area.

Without loss of generality we are at liberty to set the initial
cylinder area $A(0)=a_0=1$.  Note that this is in addition to having
set the initial cylinder length $L(0)=1$ in our choice of units above.
It is important to realise, however, that we are not restricting
ourselves to situations in which the initial area and length are
constrained relative to each other in any particular way.  Any
information about the relative values of the cylinder's area and
length has simply been lost as a consequence of making the slender
filament approximation \cite{Denn1975,Clasen2006}.

\subsection{Heterogeneous perturbations}

\begin{figure*}
	\centering
	\includegraphics[height = .90\textheight]{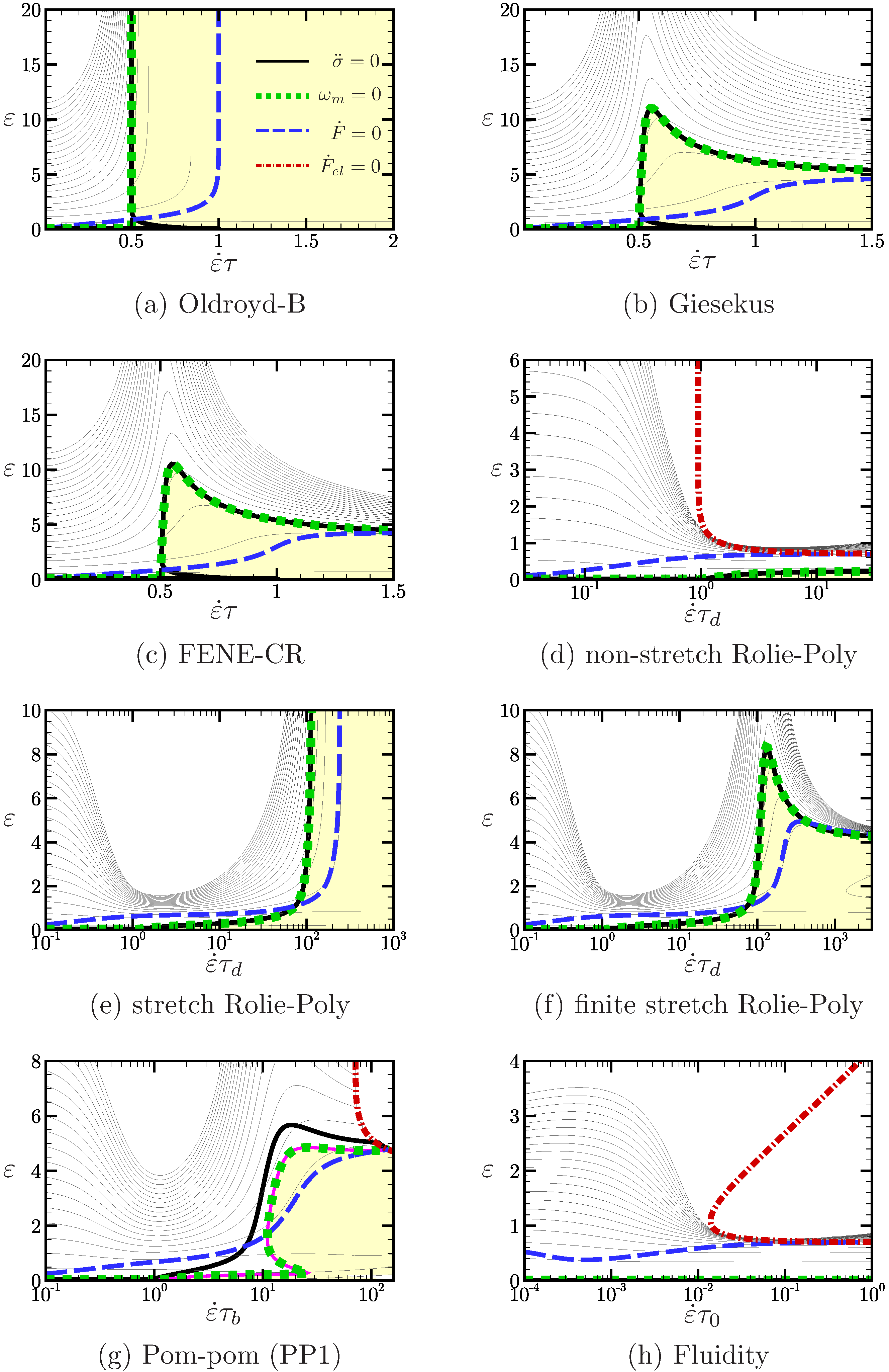}
	\caption{Numerical results for the linearised necking dynamics within
		six tensorial constitutive models. (Results for the Rolie-Poly model
		are shown separately with chain stretch disallowed (d); with chain
		stretched allowed and unrestricted (e); and with chain stretched
		allowed but restricted to be finite (f).) The thin black lines show
		contours of constant area perturbations $\delta a / \delta a_0 =
		10^{n/4}$, with $n = 1 \cdots 20$ in curves from bottom to top,
		representing the growing degree of necking at increasing strain
		$\epsilon$ upwards in any filament stretching experiment at fixed
		$\edot$. The green dotted lines shows the strain at which the
		largest eigenvalue becomes positive, in increasing strain $\epsilon$
		upwards during stretching at fixed $\edot$. (The overhang in the
		Pom-pom model gives stability-instability-stability-instability in
		the range of strain rates $10< \edot\tau < 30$.) Accordingly, the
		beige shaded area shows the window of strains over which the
		filament is stable against necking. The thick black
		solid line shows the strain at which the stress curvature
		criterion predicts the onset of necking, in increasing $\epsilon$
		at fixed $\edot$. The red dot-dashed line shows the strain at which
		the ``elastic \considere'' mode becomes unstable.  The blue
		long-dashed line shows the strain at which the original \considere
		criterion would predict the onset of necking. 
		\label{fig:LSA}}
\end{figure*}

So far we have discussed a calculation in which the filament is
assumed to remain perfectly uniform as it is stretched out, with all
the flow variables homogeneous along it. In the language of
hydrodynamic stability theory, this gives a ``base state''
$(\edotbar,a_0,\conf_0(t),\lambda_0(t),Q_0(t),\cdots)$. In contrast to
conventional stability calculations, however, this base state is
time-dependent, because of the time-evolution of the viscoelastic
conformation variables following the onset of the applied flow, giving
rise to the time-evolving tensile stress signals in
Fig.~\ref{fig:stress} (left panels).  Eventually, after several strain
units, the base state will attain a steady state with a tensile stress
on the homogeneous constitutive curve. As the calculations that follow
will show, however, the sample will in general neck significantly
before the system has a chance to attain a state of homogeneous flow
on that constitutive curve.

To study how this necking arises, we now add to the homogeneous base
state small amplitude heterogeneous perturbations, decomposed into
Fourier modes with wavevector $q$ that is reciprocal to the space
variable $u$ along the filament's transformed length:
\begin{equation}
	\begin{pmatrix} \edot(u,t)\\ a(u,t)	\\ \conf(u,t)\\
                \lambda(u,t)\\                Q(u,t)\\                \vdots\\
	\end{pmatrix}
= 
	\begin{pmatrix}
		\edotbar\\    a_0	\\		\conf_0(t)\\
                \lambda_0(t)\\                Q_0(t)\\                \vdots\\
        \end{pmatrix}
+
	\sum_q\begin{pmatrix}
		\delta\edot(t)\\ \delta a(t)	\\	\delta \conf(t)\\
                \delta \lambda(t)\\ \delta Q(t)\\       \vdots\\
	\end{pmatrix}_q \exp(iqu)
	. \label{eqn:matrixM}
\end{equation}
Note that the area perturbations $\delta a(t)$ obey $\delta
a(t)/a_0=\delta A(t)/A(t)$ where $a_0$ is constant. They thereby give
a measure, at any time $t$, of the fractional variations in cross
sectional area along the filament's length, compared to the
length-averaged cross sectional area at that time. They therefore
indicate the degree to which the filament has necked at that time.

We then substitute this expression (\ref{eqn:matrixM}) into the
governing equations (\ref{eqn:massT}) to (\ref{eqn:confT}), expand in
successive powers of the amplitude of the perturbations, and retain
only terms of first order in this amplitude. This gives linearised
equations for the dynamics of the perturbations:
\begin{equation}
\partial_t \begin{pmatrix}
		\delta\edot(t)\\ \delta a(t)	\\	\delta \conf(t)\\
                \delta \lambda(t)\\ \delta Q(t)\\       \vdots\\
	\end{pmatrix}_q 	
= \tens{M}(t)\cdot
\begin{pmatrix}
		\delta\edot(t)\\ \delta a(t)	\\	\delta \conf(t)\\
                \delta \lambda(t)\\ \delta Q(t)\\       \vdots\\
	\end{pmatrix}_q, \label{eqn:matrixMb}
\end{equation}
in which it is important to note that the stability matrix
$\tens{M}(t)$ has inherited the time-dependence of the base state
$(\edotbar,a_0,\conf_0(t),\lambda_0(t),Q_0(t),\cdots)$, upon which it
depends.

We note that the stability matrix $\tens{M}(t)$ has however no
dependence on the wavevector $q$. This stems from the fact that the
governing equations (\ref{eqn:massT}) to (\ref{eqn:confT}) are
spatially local, apart from the convective terms, which drop out at
linear order in the above expansion.  (Spatial dependence of
$\tens{M}$ would however be restored at lengthscales shorter than
those considered here by moving beyond our slender filament
assumption, or by incorporating surface tension, or by including
stress diffusion, which would cutoff any instability at short
lengthscales.) In this way, all Fourier modes $\exp(iqu)$ are
predicted to have the same dynamics.  Which mode will dominate any
necking stability in practice is therefore determined by which is
initially seeded most strongly: whether by thermal noise, slight
initial sample imperfection, or by geometrical features of the
experimental device. In a filament stretching rheometer we expect the
dominant seeding to arise from the no-slip condition that applies
where sample ends meet the rheometer plates. As noted above, this
constrains the area to remain constant at each of the sample ends as
the sample is stretched out overall, thereby initiating a single neck
in the middle of the filament.  (It is this effect that is modelled in
our nonlinear simulations by the boundary condition discussed at the
start of Sec.~\ref{sec:NL}.)

To determine whether in any filament stretching experiment the
perturbations $\delta a(t)$ in the filament's cross sectional area
will start to grow towards a necked state, and at what time during the
run they first start to do so, these linearised equations must be
integrated in time. If the stability matrix $\tens{M}$ were
time-independent, it would be trivial to establish that $\delta a$
would grow if at least one eigenvalue of $\tens{M}$ had positive real
part.  The time-dependence of $\tens{M}$ clearly complicates this,
although a good indication of whether the area fluctuations $\delta a$
will be growing at any time during a run is given by whether (at
least) one of the time-dependent eigenvalues of $\tens{M}$ has a
positive real part at that time. However the concept of a
time-dependent eigenvalue is clearly delicate. In our numerical
simulations of the six constitutive models in
Sec.~\ref{sec:results_lsa} below, therefore, we both directly
integrate the linearised equations {\em and} report the more
  indirect measure given by the sign of the real part of the
time-dependent eigenvalue of $\tens{M}$ with the largest real part.
Pleasingly, we find good agreement between the regime of strongly
growing area fluctuations $\delta a$, and of a positive eigenvalue of
$\tens{M}$.

The aim of the next section is to analytically derive fluid-universal
(model-independent) criteria for the time at which this necking
instability first sets in during any filament stretching run, which we
can then compare with our numerical simulations of the six
constitutive models.

\section{Criteria for necking}
\label{sec:criteria}

Having outlined the procedure of a linear stability analysis in
general terms, we now perform this calculation analytically in the
particular case of the simplified scalar model introduced in
Sec.~\ref{sec:toy}.  Our goal in doing so is to derive criteria for
the onset of necking that can be reported in terms of characteristic
signatures in the shapes of the material response functions (tensile
stress $\sigmae^+$, {\it etc.}) as a function of the time (or
accumulated strain) since the inception of the flow. Recall that the
original \considere criterion would predict the onset of necking to
coincide with the characteristic signature $\partial_{\epsilon}F<0$ in
the functional form of the tensile force $F$ as a function of the
accumulated strain $\epsilon$. As noted above, however, there is no
reason a priori to expect this criterion for necking in solids to
apply in complex fluids with a finite stress relaxation timescale
$\tau$.

We start by recollecting for convenience the governing equations. The
condition of mass balance gives
\be
\label{eqn:massTb}
\partials{a}{t} + (v-\edotbar u)\partials{a}{u} = -(\edot-\edotbar)a.
\ee
The condition of force balance for the transformed force
$\ftrans(t)=F(t)\exp(\edotbar t)$ gives
\be
\label{eqn:forceTb}
 0 = \partials{\ftrans}{u}=\partials{(a\sigmae)}{u},
\ee
where the tensile stress
\be
\label{eqn:stressToyb}
\sigmae=G Z + \eta\edot.
\ee
The scalar conformation variable evolves as
\be
\label{eqn:constitb}
\partials{Z}{t}+(v-\edotbar u)\partials{Z}{u}=\edot f(Z)-\frac{1}{\tau}g(Z),
\ee
We intentionally leave unspecified the forms of the loading and
relaxation functions $f(Z)$ and $g(Z)$ in order that the criteria we
derive are as fluid-universal as possible, independent of particular
constitutive choices.

Following the procedure outlined above, we consider a uniform base
state $\edotbar,a_0,Z_0(t)$, in which the viscoelastic conformation
variable $Z_0(t)$ evolves as a function of the time $t$ since the
inception of the flow in a manner prescribed by
Eqn.~\ref{eqn:constitb}, solved within the assumption of homogeneous
flow, $\partial_uZ_0=0$.  The tensile stress
$\sigmaeb^+(t)=GZ_0(t)+\eta\edotbar$ accordingly evolves (in the absence
of pathological choices for the loading and relaxation functions $f$
and $g$) towards its eventual steady state on the homogeneous
constitutive curve at the given applied strain rate $\edotbar$. The
corresponding tensile force $F_0(t)=\sigmaeb^+(t)A_0(t)$ initially
increases as a result of the rising stress, then decreases due to the
declining filament area $A_0(t)=A(0)\exp(-\edotbar t)$.

To this uniform, time-evolving base state we now add small amplitude
heterogeneous  perturbations, which are the precursor of any neck:
\begin{equation}
	\begin{pmatrix} \edot(u,t)\\ a(u,t)	\\ Z(u,t)\\
	\end{pmatrix}
= 
	\begin{pmatrix}
		\edotbar\\    a_0	\\		Z_0(t)\\
        \end{pmatrix}
+
	\sum_q\begin{pmatrix}
		\delta\edot(t)\\ \delta a(t)	\\	\delta Z(t)\\
	\end{pmatrix}_q \exp(iqu)
	. \label{eqn:matrixMc}
\end{equation}
(We have kept the $q$ subscript here as a reminder that we consider
perturbations on all spatial lengthscales. However the reader should
recall the discussion in the previous section, noting that the
stability matrix is actually $q-$indpendent and that instead the
dominant mode will be the one seeded most strongly by endplate
effects, leading to a single neck mid-filament.)
Substituting~(\ref{matrixMx}) into the governing equations
(\ref{eqn:massTb}) to (\ref{eqn:constitb}), expanding in powers of the
amplitude of the perturbations, and retaining only terms of first
order in that amplitude, gives a set of linearised equations for the
dynamics of the perturbations.

The linearised mass balance equation is
\be
\partials{\delta a_q}{t}=-\delta\edot_q.
\ee
The linearised force balance equation is
\be
0=\sigmae \delta a_q + G\delta Z_q + \eta \delta\edot_q,
\ee
and the linearised viscoelastic constitutive dynamics
\be
\partials{\delta Z_q}{t}=\delta \edot_q f(Z_0)+C\delta Z_q
\ee
in which 
\be
C=\edot f'(Z_0)-\frac{1}{\tau}g'(Z_0).
\ee
Here and throughout we use a prime to denote differentiation with
respect to a function's own argument.  Eliminating $\delta \edot_q$,
which is instantaneously slaved to the other variables by the
condition of force balance in creeping flow, gives the two-dimensional
linearised equation set:
\begin{equation}
\partial_t
        \begin{pmatrix}
		\delta a(t)	\\ \\	\delta Z(t)\\
	\end{pmatrix}_q 
= \tens{M}(t)
\cdot
        \begin{pmatrix}
		\delta a(t)	\\  \\ 	\delta Z(t)\\
	\end{pmatrix}_q. \label{eqn:2Dset}
\end{equation}
This is governed by the stability matrix
\be
\tens{M}(t)= 	\begin{pmatrix} \dfrac{\sigma_{\rm E0}}{\eta} & \dfrac{G}{\eta} \\
    & \\
                        \dfrac{-f(Z_0)\sigma_{\rm E0}}{\eta}\;\;\;\ & -\dfrac{f(Z_0)G}{\eta} +C \\
     & \\
	\end{pmatrix}.
\ee
This depends on time via the evolving viscoelastic conformation
variable $Z_0(t)$ in the base state, until that base state settles to
a steady state with the tensile stress on the stationary homogeneous
constitutive curve. As we shall see below, however, the sample in
general necks significantly before any such stationary homogeneous
state can be attained.

Denoting the trace and determinant of $\tens{M}$ by $T$ and $\Delta$
respectively, we have
\be
\label{eqn:trace}
T=\frac{1}{\eta}(\sigmaeb-fG)+O(1),
\ee
and
\be
\label{eqn:det}
\Delta=\frac{\sigmaeb C}{\eta}.
\ee
In what follows we assume the solvent viscosity $\eta$ small compared
to the scale of the viscoelastic viscosities, and accordingly ignore
the $O(1)$ term in $T$.

It is possible to show that these expressions for $T$ and $\Delta$ can
be cast in terms of more physically meaningful properties of the
underlying base state as follows:
\be
\label{eqn:traceb}
T=-\frac{1}{A(0)\eta}\EC,
\ee
and
\be
\label{eqn:detb}
\Delta=\frac{\sigmaeb}{\eta}\frac{ \ddSb}{\dSb},
\ee
where dot denotes time differentiation.  Note that the derivative
$\EC$ of the base state's tensile force with respect to strain
$\epsilon$ in Eqn.~\ref{eqn:traceb} needs careful interpretation. It
is defined by evolving the state up to some strain $\epsilon$ with the
full model dynamics, including loading by flow as encoded by $f$ and
relaxation as encoded by $g$. In the next increment of strain
$\epsilon \to \epsilon +\delta \epsilon$ over which the derivative is
taken the relaxation term $g(Z)$ is then suppressed, with only the
loading (`elastic') dynamics implemented.  Necking in the elastic
limit of viscoelastic models was also discussed in~\cite{Hassager1998}.

The time-dependent eigenvalues $\omega(t)$ of $\tens{M}$ follow as
solutions of the quadratic
\be
\omega^2-T\omega+\Delta =0.
\ee
By examining the conditions under which these can become positive (in
the sense of having a positive real part) during the course of a
filament stretching run, we find two different possible modes of
instability to necking. The first gives instability in any regime
where the determinant $\Delta <0$, and accordingly where
\be
\frac{\ddS}{\dS}<0.
\ee
(We return below to justify the fact that we have just dropped the
base state subscript, writing $\sigmaeb\to\sigmae$.) In all the
constitutive models considered in this work the tensile stress
increases monotonically towards the constitutive curve (at least
before significant necking occurs), $\dS>0$, and the criterion for the
onset of necking instability is simply
\be
\label{eqn:curvature}
\ddS<0.\;\;\;\;\;\;\;\;\;\textbf{``Stress curvature mode''}
\ee
Its eigenvalue has an amplitude set by the imposed strain rate
$\edotbar$. Once this mode of instability sets in, therefore, it gives
rise to an observable neck over the course of a few strain units.
Note that in some experiments the stress signal has been observed to
overshoot on its approach to the constitutive curve, giving $\dS<0$
after the overshoot~\cite{Rasmussen2005}. The implications of this for
necking will be considered in future work.

The second mode gives instability in any regime where the trace $T>0$,
and so where
\be
\label{eqn:EC}
\EC<0.\;\;\;\textbf{``Elastic \considere mode''}
\ee
Its eigenvalue has an amplitude $O(G/\eta)$, which is large for the
small solvent viscosities relevant to the highly viscoelastic
materials considered here. If this mode becomes unstable during any
filament stretching run, it causes a neck to then develop very quickly
compared to the rate of the imposed flow.

Note that we have dropped the base-state subscript ``0'' in writing
(\ref{eqn:curvature}) and~(\ref{eqn:EC}). This is justified because
the base-state properties contained in these expressions -- in
particular the stress curvature $\ddS$ in (\ref{eqn:curvature}) --
also correspond to the full experimentally measured properties at
least as long as the sample remains uniform, and therefore up to the
point that significant necking occurs.  Accordingly, the
experimentally measured signals can be used in these expressions in
order to predict the onset of necking.

The stress signal $\sigmae^+$ in (\ref{eqn:curvature}) is widely
reported in the experimental literature. Whether the ``elastic''
derivative of the force $\EC$ in (\ref{eqn:EC}) has any easily
measurable counterpart is an open question.  In any case, it is
crucial to note that the condition $\EC=0$ does {\em not} correspond
to the condition $\partial_\epsilon F=0$. This new ``elastic''
\considere criterion is therefore {\em not} (in general) the same as
the original \considere criterion, which predicts necking instability
when $\partial_\epsilon F<0$. Indeed our numerical results below will
show that the two conditions predict the onset of instability at
different strains over most of the range of imposed strain rates
$\edot$. They do, however coincide as $\edot\tau\to\infty$. This is to
be expected, because in this limit a fluid's relaxation dynamics, as
encoded by $g/\tau$ here, become unimportant compared to the loading
by flow, as encoded by $\edot f$.

Note that we chose to express the stress curvature criterion in terms
of time derivatives of the stress, and the elastic \considere
criterion in terms of a strain (elastic) derivative of the force.
However both criteria can equally be expressed in either time or
strain derivatives, because $\partial_t=\edot \partial_\epsilon$ for
the constant $\edot$ protocol of interest here. We chose our notation
simply because the stress curvature criterion corresponds to a rather
gradual, liquid-like mode of instability, for which time seems the
more the natural independent variable. In contrast, the elastic
\considere criterion pertains to a solid-like mode, for which strain
seems more the natural variable.

Because these criteria for necking have been derived within a highly
generalised constitutive model, independent of any particular
constitutive choices for the functions $f$ and $g$, we suggest that
they should apply rather widely among materials that show a basic
competition between elastic loading and viscoelastic (or plastic)
relaxation. Pleasingly, our numerical results in
Sec.~\ref{sec:results_lsa} will show that they hold either exactly or
to a good level of approximation in all of the six tensorial
constitutive models considered in this work.

Indeed, making the concrete choices $f=3 + 2Z - \beta Z^2$ and $g= Z +
\alpha Z^2$ for the loading and relaxation functions in this scalar
model gives an excellent approximation to the necking predictions of
the full tensorial Oldroyd B model ($\alpha=\beta=0$), the Giesekus
and FENE-CR models ($\alpha\neq 0,\beta=0$) and the Rolie-Poly model
without chain stretch ($\alpha=0,\beta\neq 0$). We verified this by
comparing numerical results for the scalar model with these parameter
choices with their counterparts in the full tensorial models, finding
excellent agreement (Appendix~\ref{app:Toy-Model}).

\section{Numerical results: linear stability}
\label{sec:results_lsa}

Having derived criteria for the onset of necking, we now present
numerical results to demonstrate their validity within six widely used
tensorial constitutive models.  See Fig.~\ref{fig:LSA}. Each panel of
this figure contains results presented in the plane ($\epsilon,\edot$)
of strain and strain rate (with the latter expressed in units of
the characteristic relaxation time in any model).  Any vertical line
from bottom to top in this plane tracks a single filament stretching
run performed at constant strain rate $\edot$, with the strain
$\epsilon$ increasing upwards in the plane as the filament is
stretched out.  (Note that in previous sections we used $\edotbar$ to
denote the strain rate applied globally to the sample as a whole, and
$\edot(z,t)$ the strain rate that varies locally along the filament as
the sample necks. In the results sections that follow, we often use
$\edot$ to denote the globally applied strain rate for notational
convenience.)

The thin solid black lines show contours of constant area perturbation
$\delta a(t)$.  (Recall the note after Eqn~\ref{eqn:matrixM}
  regarding the definition of $\delta a$.) Each successive contour
crossing (as $\epsilon$ increases upwards at fixed $\edot$)
corresponding to an increase of factor $10^{1/4}$ in the area
perturbation, such that the $nth$ contour represents a degree of
necking $\delta a / \delta a_0 = 10^{n/4}$, where $\delta a_0$ is the
initial small seeding provided at the start of the run. The more
densely clustered the contour lines vertically at any fixed $\edot$,
therefore, the faster necking occurs in a filament stretching run at
that strain rate. We have shown only the first $20$ contour lines,
assuming that the sample will have failed altogether by this time.
Indeed, although the absolute necking amplitudes associated with these
contours are arbitrary at the level of this linear analysis, a good
indication of the dependence of the strain at which the sample is
likely to finally fail on the imposed strain rate is given by focusing
on one representative contour.

Also included in each panel is a green dotted line showing the strain
$\epsilon$ at which the largest eigenvalue $\omega_m$ first becomes
positive (in the sense of having a positive real part) in any
filament stretching run at a given strain rate $\edot$, signifying the
onset of instability to necking.  (The slight overhang in the Pom-pom
model gives stability-instability-stability-instability in the range
of strain rates $10< \edot\tau < 30$.) Accordingly, the beige shading
shows the window of strains in which the sample is stable against
necking at any given imposed strain rate. As can be seen, the onset of
a positive eigenvalue agrees convincingly with the onset of strong
exponential growth in necking depicted by the contour lines.  (Some
transient growth is seen before the eigenvalue becomes positive, but
of an amplitude unlikely to be seen experimentally.)

The thick black solid line in each panel depicts the strain at which
the stress curvature criterion predicts the onset of necking, as the
strain increases in any filament stretching run at fixed $\edot$. For
strains below this line, the stress is an upwardly curving function of
time (or accumulated strain). For strains above it, the stress curves
downward in time as it tends finally towards the constitutive curve.
(Recall Fig.~\ref{fig:stress}.) As can be seen, this stress curvature
criterion exactly predicts the onset of necking ({\it i.e.,} the
change in sign of the eigenvalue) in the Oldroyd B, Giesekus, FENE-CR,
and Rolie-Poly models, apart from a slight discrepancy at small
strains 
~\footnote{This arises because $GW_{zz}$ and $\sigmae$ do not
  coincide in this regime: instead $W_{xx}$ and $W_{zz}$ are of
  comparable size and both contribute to $\sigmae$.  Recall that our
  derivation of the criterion above assumed $W_{zz}\equiv Z$ to be
  large compared to all other components of $\tens{W}$.}. 
  It also performs qualitatively well in the Pom-pom model, with some
quantitative discrepancy 
~\footnote{This quantitative discrepancy
  arises because in the Pom-pom model the viscoelastic tensile stress
  depends not only on $\tens{W}$ but also on the auxiliary variable
  $\lambda$.  Our derivation of onset criterion assumed the
  viscoelastic stress to be given simply by $GW_{zz}$.}.  
  In the fluidity model the curvature criterion is negative from the inception
of flow for all strain rates.

In the Oldroyd B, Giesekus, FENE-CR and Rolie-Poly model (with chain
stretch), the stress curvature mode of necking instability just
discussed is the only one that arises. As noted above, it leads to
relatively gentle necking with an eigenvalue of characteristic
amplitude set by $\edot$.  In stark contrast, at high strain rates in
the Roly-Poly model without chain stretch, and in the Pom-pom and
fluidity models, the ``elastic \considere'' mode also arises. The
strain at which it sets in is indicated by the red dot-dashed lines.
Once active, it causes very fast necking, with an eigenvalue
$O(G/\eta)$: the contours of $\delta a$ then become almost too
narrowly spaced to be separately discerned on the strain scale of the
plots in Fig.~\ref{fig:LSA}, and the sample fails catastrophically
quickly.

This difference in severity between the stress curvature and elastic
\considere modes leads to radically different necking dynamics at low
and high strain rates in these three models
(Figs.~\ref{fig:LSA}d,g,h). At low strain rates, necking occurs
gradually via the stress curvature mode. At high strain rates the
sample violently fails via the elastic \considere mode.  However in
any case where the elastic \considere mode does become unstable, it
does so only after the stress curvature mode has already caused some
necking.  In this sense, it can be seen as a rapid amplifier of an
already existing, but much gentler instability. (In the fluidity model
a return to elastic-\considere stability is seen at larger strains at
high strain rates, but this is of no relevance because the sample will
have already entirely failed by that point.)

Having discussed the general features of the plots in
Fig.~\ref{fig:LSA}, we now describe in more detail the necking
dynamics of each constitutive model in turn, starting with the Oldroyd
B model in panel a). Here the dominant contribution $GW_{zz}$ to the
tensile stress $\sigmae^+$ has a time-dependence of shape
$1+\exp\left[(2\edot-1/\tau)t\right]$.  For imposed strain rates
$\edot<1/2\tau$, therefore, the stress curves downwards as a function
of time, and a neck starts to form, right from the inception of the
flow. In contrast, for imposed strain rates $\edot>1/2\tau$ the stress
signal curves upwards in time indefinitely. This is a result of the
fact that the molecular dumbbells that this model describes can
stretch out indefinitely with flow, consistent with the absence at
these strain rates of a steady state in the constitutive curve of
Fig.~\ref{fig:constitutive}a).  Although this well known `extensional
catastrophe' indefinitely stabilises the filament against necking, it
is an unphysical feature of the Oldroyd B model.  We note that the
predictions of Fig.~\ref{fig:LSA}a) agree with earlier work in
\cite{Olagunju1999}. 

The FENE-CR and Giesekus models each contain nonlinearities that
regularise this extensional catastrophe, cutting off the indefinite
dumbbell stretching and restoring a finite stress even at strain rates
$\edot > 1/2\tau$ in the constitutive curves of
Fig.~\ref{fig:constitutive}b,c).  (Both models reduce back to Oldroyd
B when those nonlinearities are removed.) The Giesekus model, for
example, includes an additional term in the relaxation dynamics,
$Z/\tau\to (Z+\alpha Z^2)/\tau$, with $\alpha$ small, recovering
Oldroyd B when $\alpha=0$. These nonlinearities manifest themselves
only once the dumbbells become strongly stretched, such that $\alpha Z
= O(1)$.  Because of this, for strain rates $\edot < 1/2\tau$ the
dynamics of Giesekus and FENE-CR are essentially the same as Oldroyd
B, with a downwardly curving stress $\sigmae^+(t)$ and instability to
necking.  Similarly at higher strain rates $\edot > 1/2\tau$, and for
strains less than around $5-10$, before strong chain stretch arises,
Giesekus and FENE-CR display an upwardly curving $\sigmae^+(t)$ and
stability against necking, as in Oldroyd B.  Compare panels b,c) and
a) in Fig.~\ref{fig:LSA}.

In contrast, beyond a typical strain $5-10$ for strain rates $\edot >
1/2\tau$, the chain stretch becomes significant and the nonlinearities
of Giesekus and FENE-CR become important, departing from the dynamics
of Oldroyd B.  In particular the chain stretch saturates, the upward
curvature in the stress signal is halted, and the stress displays an
inflexion point $\ddS=0$, beyond which the stress curves downwards as
a function of time (or accumulated strain) as it tends towards the
constitutive curve (which it in general however will not reach before
significant necking occurs). The filament then, indeed, becomes
unstable to necking, as seen by the closely gathered contours beyond
the green dotted line in Figs.~\ref{fig:LSA}b,c)

The three models discussed so far provide phenomenological
descriptions of the rheology of polymer solutions. We now turn to the
microscopically motivated Rolie-Poly (RP) model, which describes more
concentrated solutions and melts of entangled linear polymers.  It
contains the basic dynamical processes of reptation, in which the
orientation of a test chain relaxes on a timescale $\taud$; chain
stretch, which relaxes on a much faster timescales $\taus$; and
convective constraint release (CCR), in which the relaxation of chain
stretch releases entanglement points and so also allows some
relaxation of orientation.

Before describing the necking predictions of the RP model, we recall
its extensional constitutive curves in
Figs.~\ref{fig:constitutive}d,e,f).  Setting $\taus\to 0$ gives the
non-stretching Rolie-Poly (nRP) model, for which the constitutive
curve (Fig.~\ref{fig:constitutive}d) has a regime of linear response
at low strain rates, then a plateau for strain rates $\edot > 1/\taud$
where the chain orientation saturates.  Restoring a finite $\taus$
gives the stretching Rolie-Poly (sRP). For strain rates $\edot\ll
1/\taus$ its constitutive curve (Fig.~\ref{fig:constitutive}e) is
essentially the same as that of the nRP model, with minimal chain
stretch. For $\edot=O(1/\taus)$ and above, however, significant chain
stretch develops and the stress correspondingly increases.  Indeed in
the raw form of the sRP model the stress diverges as $\edot\to
1/\taus$, in direct counterpart to the stretch catastrophe of the
Oldroyd B model.  Including `FENE' terms to give the finite-stretch
Rolie-Poly (fsRP) model regularises this, as in moving from Oldroyd B
to FENE-CR, eliminating the possibility of indefinite chain stretch
and restoring a finite stress at all strain rates
(Fig.~\ref{fig:constitutive}f).

We now address the necking predictions of the Rolie-Poly model,
starting with the version without chain stretch (the nRP model).  See
Fig.~\ref{fig:LSA}d. Here the stress curvature criterion predicts
necking instability right from the inception of the flow for strain
rates $\edot<1/\taud$, and after a only a modest accumulated strain
for $\edot > 1/\taud$.  Furthermore, in this regime $\edot > 1/\taud$
the elastic \considere mode also sets in after an accumulated strain
$\epsilon=O(1)$, causing violently fast necking once it does so.  Just
as in Oldroyd B the indefinite chain stretch and associated divergence
in tensile stress indefinitely stabilised a filament against necking,
so conversely the absence of chain stretch and the associated
saturation in the stress with increasing strain rate cause
dramatically fast necking instability via the elastic \considere mode
in the nRP model for strain rates $\edot>1/\taud$.  Intuitively, as
the sample starts to neck and the strain rate in the necking region
increases, the polymer is unable to provide any counterbalancing
stress to restabilise the flow and the sample quickly fails.

In order for the elastic \considere mode of instability to arise in
any given constitutive model, the `loading' terms of that model, as
represented by $\edot f$ in the toy scalar version discussed above,
must be sufficiently nonlinear that the `elastic' force derivative
$\EC$, which is set by $-(\sigmae^+-Gf)$, can become negative. For
most models of polymeric flows that we have considered, in most flow
regimes, this does not happen. In the nRP model, however, the loading
dynamics has a (somewhat counterintuitive) negative contribution: in
its toy representation, the loading term $f=3+2Z-\tfrac{2}{3}Z^2$. The
origin of the negative contribution lies in the assumption of
instantaneous chain stretch relaxation $\taus\to 0$: what would be a
chain stretch relaxation process with a finite associated timescale
$\taus$ in the full sRP model is instead assumed to be infinitely fast
in nRP, such that the chain stretch relaxes as quickly as it builds up
with rate $\edot$ in the flow. This leads to the apparent `negative
loading' term $-\tfrac{2}{3}\edot Z^2$: it is actually a relaxation
process, but occurring at the rate the sample is `loaded' by strain.
At high strain rates, this causes $f$ to eventually become negative,
sending $-(\sigmae^+-Gf$) and $\EC$ negative, giving elastic \considere
instability.

The stretching Rolie-Poly (sRP) model has a finite chain stretch
relaxation timescale $\taus$, which allows chain stretch to develop
for strain rates $O(1/\taus)$ and above. Indeed, as noted above, in
its raw form the sRP model displays an extensional catastrophe for
strain rates $\edot>1/\taus$, with indefinitely increasing chain
stretch and tensile stress. As in Oldroyd B, this indefinitely
stabilises the filament against necking at imposed strain rates $\edot
> 1/\taus$: in Fig.~\ref{fig:LSA}e), no contour lines are crossed as
the strain $\epsilon$ increases at these large strain rates (apart
from mild transient growth, which quickly saturates).  The
introduction of ``FENE'' terms into the sRP model regularises this
indefinite molecular stretching, rendering it finite and restoring
necking instability at all strain rates (Fig.~\ref{fig:LSA}f).

The necking dynamics of the full Rolie-Poly model with finite chain
stretch (Fig.~\ref{fig:LSA}f) therefore shows four distinct regimes,
which can be summarised as follows.  (I) For imposed strain rates $\edot <
1/\taud$, gentle necking occurs right from the inception of the flow.
(II) For strain rates in the regime $1/\taud < \edot < 1/\taus$ the
saturation of chain orientation associated with the flat region in the
underlying constitutive curve places strong limitations on the tensile
stress that the polymer can provide in any developing neck, giving
more violent necking. This is a vestige in the fsRP model of the
elastic \considere mode seen in the nRP model, although true elastic
\considere doesn't arise with chain stretch. (III) For $\edot =O(1/\taus)$
the rapid rise in chain stretch rate strongly mitigates necking,
deferring it to a Hencky strain $\epsilon\approx 8$ for the parameter
values considered here.  Finally (IV) for $\edot > 1/\taus$ the chain
stretch saturates, leading to more rapid necking.

These four regimes of necking dynamics have strong signatures in the
underlying homogeneous constitutive curve
(Fig.~\ref{fig:constitutive}f): the flatter this curve at any given
strain rate, the more violent the necking at that imposed strain rate.
Perhaps surprisingly, this is true even though the system in general
does not have a chance to finally attain a state of steady homogeneous
flow on the constitutive curve before it necks. This correspondence
can, however, be motivated by a `back of the envelope' calculation for
a filament that does, in a thought experiment at least, manage to
remain uniform long enough to attain a state of steady flow on its
homogeneous constitutive curve, before starting to neck.  A linear
stability analysis for the dynamics of heterogeneous perturbations
about this state of uniform steady flow then gives the linearised
condition of mass balance
\be
\label{eqn:boeMass}
\dot{\delta a}_q=-\delta \edot_q.
\ee
The condition of force balance for a filament in a state of steady
flow on its constitutive curve $\sigmae(\edot)$ gives $\partials{[a
  \sigmae(\edot)]}{u}=0$, which when linearised gives
\be
\label{eqn:boeForce}
0 = \sigmae(\edot) \delta a_q + \sigmae'(\edot) \delta \edot_q,
\ee
where prime denotes derivative with respect to the function's own
argument. Combining these gives
\be
\frac{\dot{\delta a}_q}{\delta a_q}=\frac{\sigmae(\edot)}{\sigmae'(\edot)},
\ee
with a growth rate given by $\sigmae(\edot)/\sigmae'(\edot)$. The
necking at any given imposed strain rate is therefore faster the
flatter the underlying constitutive curve at that strain rate, as seen
in our numerical results. This can be understood physically as
follows.  Eqn.~\ref{eqn:boeMass} states that in any region of a
filament where the strain rate increases, the area decreases faster by
mass conservation.  Eqn.~\ref{eqn:boeForce} states that in any such
region where the area decreases, the strain rate must increase to
provide an enhanced stress to maintain a uniform force along the
filament. This gives a positive feedback loop and necking instability
for any constitutive curve with a positive slope $\sigmae'(\edot)>0$.
The instability is furthermore faster the flatter the constitutive
curve, because the strain rate must increase more quickly to provide
an enhanced stress to maintain uniform force.  Interestingly, a
negative constitutive slope $\sigmae'(\edot)<0$ is predicted to confer
stability against necking, and we consider the implications of this
more fully in~\cite{stress}.


The Pom-pom model of long chain branched polymers shows similar
necking dynamics (Fig.~\ref{fig:LSA}g) to those just discussed for
linear polymers. In particular we again see four distinct regimes,
with (I): relatively gentle necking at low flow rates, (II): more
rapid necking at moderate flow rates associated with the saturation in
chain backbone orientation, (III): a degree of stabilisation against
necking provided by developing backbone stretch, and finally (IV):
more violent necking at higher strain rates due to a saturation in the
degree of backbone stretch.

Two important differences between the Rolie-Poly model of linear
polymers and the Pom-pom model of branched polymers should however be
noted.  The first is that in the Pom-pom model the stress curvature
criterion (shown by the thick black line in Fig.~\ref{fig:LSA}g),
provides only a qualitative fit to the onset of necking (as
characterised by a positive eigenvalue, shown by the thick dotted
green line). This is because the expression for the polymer stress in
the Pom-pom model contains multiplicative contributions from the chain
orientation and chain stretch, each with its own dynamical evolution.
(In contrast, the Rolie-Poly has just a single dynamical factor
$W_{zz}-W_{xx}\equiv Z$ in the stress, as does the scalar model in
which we derived the necking criteria.)  As a result, the inflexion
point $\ddS=0$ of the tensile stress provides only an approximation to
the onset of necking in the Pom-pom model.  Instead, it is the
inflexion point in the component $W_{zz}$ of molecular orientation
that actually predicts onset.  However this quantity cannot be
accessed experimentally by measuring the tensile stress signal,
because it appears in the expression for $\sigmae^+$ prefactored by the
(also) time-evolving chain stretch.

\begin{figure*}
	\centering
	\includegraphics[width=0.95\textwidth ]{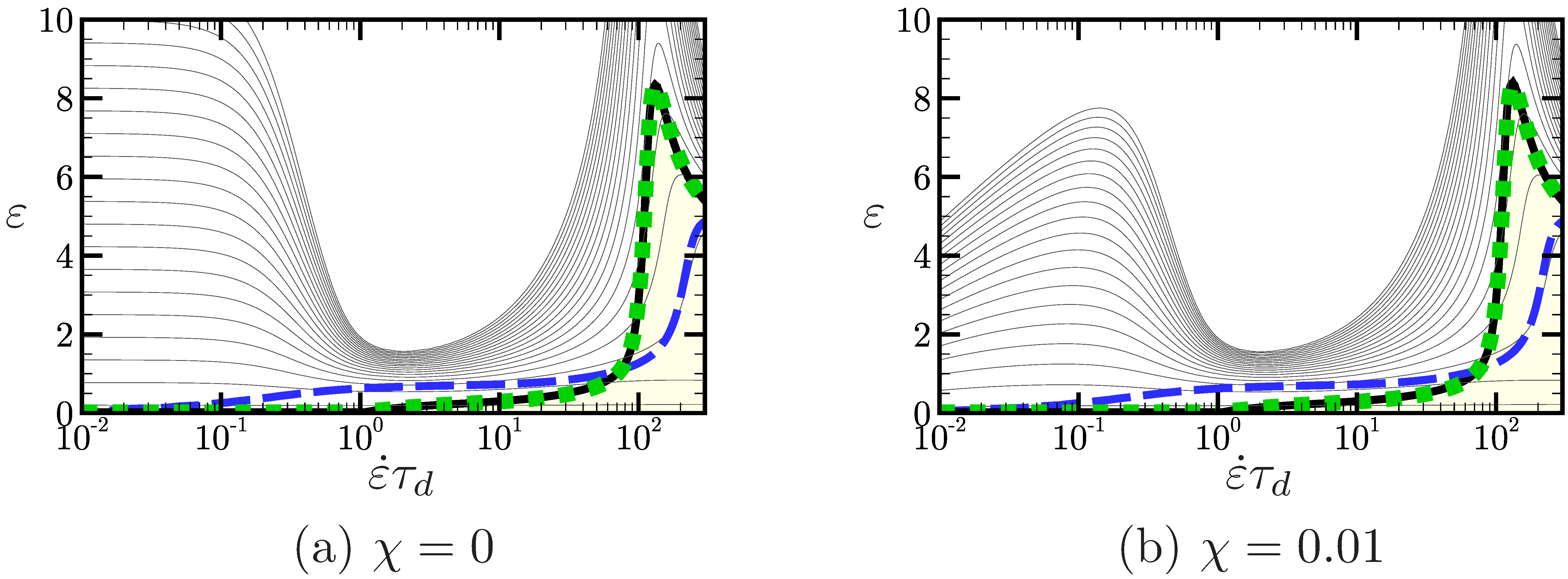}
	\caption{Linearised necking dynamics in the Rolie-Poly model
		with finite chain stretch without surface tension (a) and
		with surface tension (b). Values of all other model
		parameters are the same as in Fig.~\ref{fig:LSA}f).
		\label{fig:surface-tension}}
\end{figure*}

The second difference is the appearance at high strain rates
$\edot>1/\taus$ in the Pom-pom model of the elastic \considere mode,
which was absent in the finite stretch Rolie-Poly model. This arises
because the Pom-pom model invokes a hard cutoff for the growth in
chain backbone stretch, to model its entropic limitation by the
branching priority, $q$ \cite{Blackwell2000,McLeish1998}: the stretch
is assumed to evolve only until it is equal to $q$ and is held
constant thereafter. This sharp saturation effectively removes the
polymer's ability to provide any counterbalancing additional stress in
any thinning region of the filament, leading to violent failure.  This
upper bound on the stretch imposed in the original form of the Pom-pom
model \cite{McLeish1998,Blackwell2000} is removed in some subsequent
versions of the model (e.g.~\cite{Verbeeten2004}), causing
significantly different predictions for the necking dynamics.  We
comment further on the comparative necking dynamics predicted by these
different assumed forms of backbone stretch dynamics in
Sec.~\ref{sec:NL} below.


We consider finally the necking dynamics of soft glassy materials
(foams, emulsions, dense colloids, microgels, {\it etc.})
\cite{Sollich1996,Cates2003}. These widely exhibit rheological
ageing, in which (in the absence of an applied flow) a sample
becomes progressively more solid-like and less liquid-like as a
function of the time elapsed since it was prepared (for example by
being freshly loaded into a rheometer and presheared). The fluidity
model that we adopt captures ageing, predicting the stress relaxation
timescale $\tau$ to increase linearly (in the absence of flow) as a
function of the time since sample preparation.  Here we consider a
sample of age $\tw=1000\tau_0=1000$ (in our units) at the time the
filament stretching run commences.

As seen in Fig.~\ref{fig:LSA}h), the sample necks by qualitatively
different modes according to whether the imposed strain rate is fast
or slow compared to the inverse sample age. In fast stretching $\edot
\gg 1/\tw$, it fails violently via the elastic \considere mode. In
slower stretching, $\edot \ll 1/\tw$, it fails more gradually, via a
mode that is closely related to the stress curvature mode derived
above in the scalar model: the stress curvature criterion is negative
from the inception of flow for all $\edot$. See \cite{Hoyle2015} for
further details.  Put differently, for a fixed flow rate $\edot$ a
young sample $\tw\ll 1/\edot$ fails gradually, while an old sample
$\tw\gg 1/\edot$ will fail more dramatically via the elastic
\considere mode.  This is the first instance, of which we are aware,
in which a sample is predicted to fail by qualitatively different
modes of necking instability according simply to its own age. The same
picture also holds within the more sophisticated soft glassy rheology
model~\cite{Sollich1996, Cates2003, Hoyle2015}.

In Fig.~\ref{fig:LSA}, then, we have shown our stress curvature and
elastic \considere criteria to perform well in predicting the onset of
necking in six widely used tensorial constitutive models of linear
polymer solutions and melts, wormlike micelles, branched polymers, and
soft glassy materials. They further perform well in distinguishing
regions of relatively gradual necking, caused by the stress curvature
mode, from the much more dramatic failure, caused by the elastic
\considere mode. 

Finally, in each panel of Fig.~\ref{fig:LSA} we also mark as a blue
long-dashed line the strain $\epsilon$ at which the tensile force
overshoot occurs, in any filament stretching experiment at a fixed
strain rate $\edot$. For strains below this line, the force increases
with strain.  For strains above it, the force decreases and the
original \considere predicts instability to necking. As can be seen,
this original \considere criterion performs poorly in predicting the
necking dynamics. For example, in Oldroyd B (which is however
pathological in extension for the reasons discussed above) it fails to
predict the strain rate $\edot=1/2\tau$ below which necking occurs and
above which the filament is stable. (It instead incorrectly predicts
this threshold to be $\edot=1/\tau$.) In the Giesekus, FENE-CR,
finite-stretch Rolie-Poly and Pom-pom models it fails to predict the
pronounced upward `nose' shaped region of strains in which the sample
is initially stable against necking, before necking sets in for larger
strains. Finally in the non-stretch Rolie-Poly, Pom-pom and fluidity
models the \considere criterion does nothing to distinguish between
the slow, gradual necking that arises via the stress-curvature mode at
low strain rates, and the much more dramatic failure due to the
elastic \considere mode at higher strain rates. As noted above,
however, the original \considere criterion converges towards our
elastic \considere criterion (where present) in the limit of high
strain rates $\edot\tau\to\infty$, consistent with the fact that a
complex fluid displays essentially solid-like response in this limit.

\subsubsection*{Effects of surface tension}
\label{sec:surface-tension}

So far, we have ignored the effects of surface tension. This is
expected to be a good approximation in the initial stages of necking
for most filament stretching experiments, because the samples used are
typically highly viscoelastic and of sufficiently large radius that
bulk stresses dominate surface ones.  (Inevitably, surface tension
must play some role during the final stages of pinchoff once the neck
becomes very thin. However our slender filament calculations are not
capable of capturing the details of pinchoff, and we do not consider
this here.)

To quantify any effects of surface tension on the initial onset of
necking, we recognise that its dominant effect is to give an
additional contribution to the tensile stress of
Eqn.~\ref{eqn:tstress} \cite{Entov1997}, such that now

\be
\label{eqn:tstress_surface}
\sigmae = G\left(S_{zz} - S_{xx} \right) + 3\eta\dot\varepsilon + \frac{\chi}{r}.
\ee
Here $\chi$ is the coefficient of surface tension between the filament
and surrounding fluid (usually air), and $r$ is the time-evolving
radius of the filament. (Here we use the leading order description of
the mean curvature of the filament in the final term.  The full
expression~\cite{Eggers2008} would introduce a $q-$dependence for the
growth rate. However we do not expect this distinction to be important
given our focus here on highly viscoelastic materials.)

In Fig.~\ref{fig:surface-tension} we compare linear stability results
for the Rolie-Poly model without surface tension (left panel) with
those in a calculation where surface tension is now included. As can be
seen, surface tension affects the onset of necking only at very slow
flow rates. This is to be expected: in this regime the bulk
viscoelastic stresses are relatively modest, and furthermore the
sample survives to large strains before necking. This combination of a
highly stretched sample of small radius and low bulk stresses means
that the term $\chi/r$ is no longer small in comparison to the bulk
viscoelastic stress $G\left(S_{zz} - S_{xx} \right)$. 

We have checked that this effect of surface tension is essentially the
same across all the models considered, affecting the onset of necking
only in the regime of very slow flow rates. In view of this, we
continue to neglect surface tension throughout the rest of the
manuscript. 

\section{Non-linear simulations}
\label{sec:NL}

To examine the necking dynamics once the amplitude of heterogeneity
has grown sufficiently that the linear calculations described in the
previous section no longer provide a good approximation, the full
nonlinear slender filament equations were numerically evolved. It is
to the results of these nonlinear calculations that we now turn.

Recall that in the linear analysis we considered the dynamics of
harmonic modes with wavelengths commensurate with the filament length.
We thereby implicitly adopted periodic boundary conditions,
effectively taking our filament to correspond to a torus being
stretched.  In reality, however, the filament has finite length and
makes contact with a rheometer plate at each of its ends. The no-slip
boundary condition that must be obeyed at the plates inevitably leads,
during filament stretching, to a shear component in the flow field in
the transition zone between the fluid that contacts the plates and the
region of pure extensional stretching further from the plates.

This more complicated flow near each plate cannot be properly
implemented at the level of our slender filament calculation, which
only accounts for pure extensional flow. However a reasonable mimic of
the no-slip condition can be achieved by employing the method
suggested by \citet{Stokes2000}, whereby the background solvent
viscosity is taken to diverge near the plates, thereby acting to `pin'
the fluid to the plates.  The form of the divergence was derived, for
a Newtonian fluid, by asymptotically matching a lubrication solution
near each plate with a slender filament approximation far from the
plates, and is as follows:
\be
\eta(u) = \eta 
\left[ 
1 + \frac{1}{32}\left( \frac{r_{\scalebox{0.6}{$0$}} e^{-\bar\epsilon}}{u} \right)^2 + \frac{1}{32}\left( \frac{r_{\scalebox{0.6}{$0$}} e^{-\bar\epsilon}}{1 - u} \right)^2 \right],
\ee
where $r_0$ is the initial radius of the sample. (In principle this
form should be rederived for each given constitutive equation in turn,
but that task would be prohibitively complicated and we do not tackle
it here. We assume instead that this form provides a reasonable model
of the essential physics for non-Newtonian fluids too.) Imposing this
boundary condition sets the strain rate at each end-plate to zero. In
this way the filament area remains pinned to its initial value at each
plate, and no viscoelastic stress develops there. The fluid
velocity exactly follows that of the plates.

\begin{figure}[t]
	\centering
\includegraphics[width=0.45\textwidth]{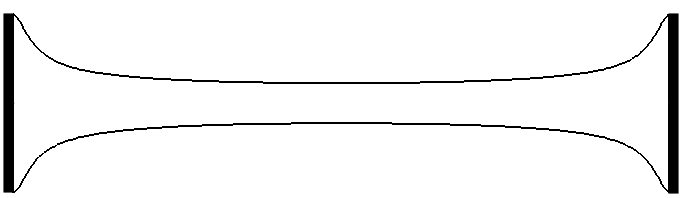}
	\caption{Effect of the no-slip boundary condition at the rheometer plates in seeding heterogeneity even before any true material instability causes significant necking. Outline shows a purely Newtonian fluid stretched to a Hencky strain $\epsilon=2.0$, within our 1D slender filament approach.
          \label{fig:BC}}
\end{figure}

Even in the absence of a true necking instability, this boundary
condition will lead to the development of some spatial heterogeneity
during filament stretching: the region of fluid near each plate is
essentially prevented from stretching properly, so the central region of
the filament thins proportionately more quickly (Fig~\ref{fig:BC}).

In view of this, one might in fact question the basis of our linear
stability calculations in Secs.~\ref{sec:LSA}
to~\ref{sec:results_lsa}, which assumed the filament to stretch in a
purely uniform way before any true necking instability arises.
However the heterogeneity arising from the no-slip condition tends to
be localised near the plates, with the central region remaining
relatively uniform until necking arises. The criteria of
Sec.~\ref{sec:criteria} should therefore still be expected to hold,
provided they are calculated using the material functions in the
central region of the filament, well away from the plates.
Nonetheless, it is reasonable to assume that the dominant source of
any initial heterogeneity that later becomes exponentially amplified
by the necking instability will be this endplate effect.

With these boundary conditions, we numerically evolve the nonlinear
slender filament equations~\ref{eqn:massT} to~\ref{eqn:confT}.  The
fact that these are already expressed in the coextending frame removes
any need for remeshing the numerical grid as the filament stretches.
Therefore we discretize the equations on a fixed mesh of grid points
and time-step the equations using an explicit Euler algorithm for the
spatially local terms and 1st  order upwinding for
the convective terms \cite{Press2007}, with results checked against a 3rd order upwinding scheme.
Care is taken to ensure convergence of the results with respect to
increasing the number of grid points $N$ and decreasing the timestep
$\Delta t$.  Generally $N=1500$ is sufficient, although in some cases
$N=2000$ is required. A timestep $\Delta=1/2N^2$ is usually adequate,
but in the case of violent necking we employ an adaptive time stepping
algorithm in which the values of the stress and area after a single
time-step $\Delta t$ are compared to the corresponding values after
two half timesteps $\Delta t/2$. If the difference between these
values exceeds a certain tolerance, $\Delta t$ is halved and the
procedure repeated until the tolerance met.

With these remarks in mind, we now present our results.  The nonlinear
necking dynamics of soft glassy materials were reported previously in
Ref.~\cite{Hoyle2015}, so we focus here on polymeric fluids. We
consider both entangled linear polymers (and wormlike micelles),
studied using the Rolie-Poly model; and entangled branched polymers,
studied using the Pom-pom model.

\subsection{Entangled linear polymers}

In this section we consider entangled linear polymeric fluids
(polymers and wormlike micelles) described by the Rolie-Poly model
with finite chain stretch. We study a highly entangled sample with
entanglement number $Z=40$, which gives a chain stretch relaxation
time $\taus=\taud/3Z=0.00833$ in our units. The values of the other
model parameters are detailed in App.~\ref{app:RP}.

\begin{figure*}[t]
	\centering
	\includegraphics[height=.7\textheight]{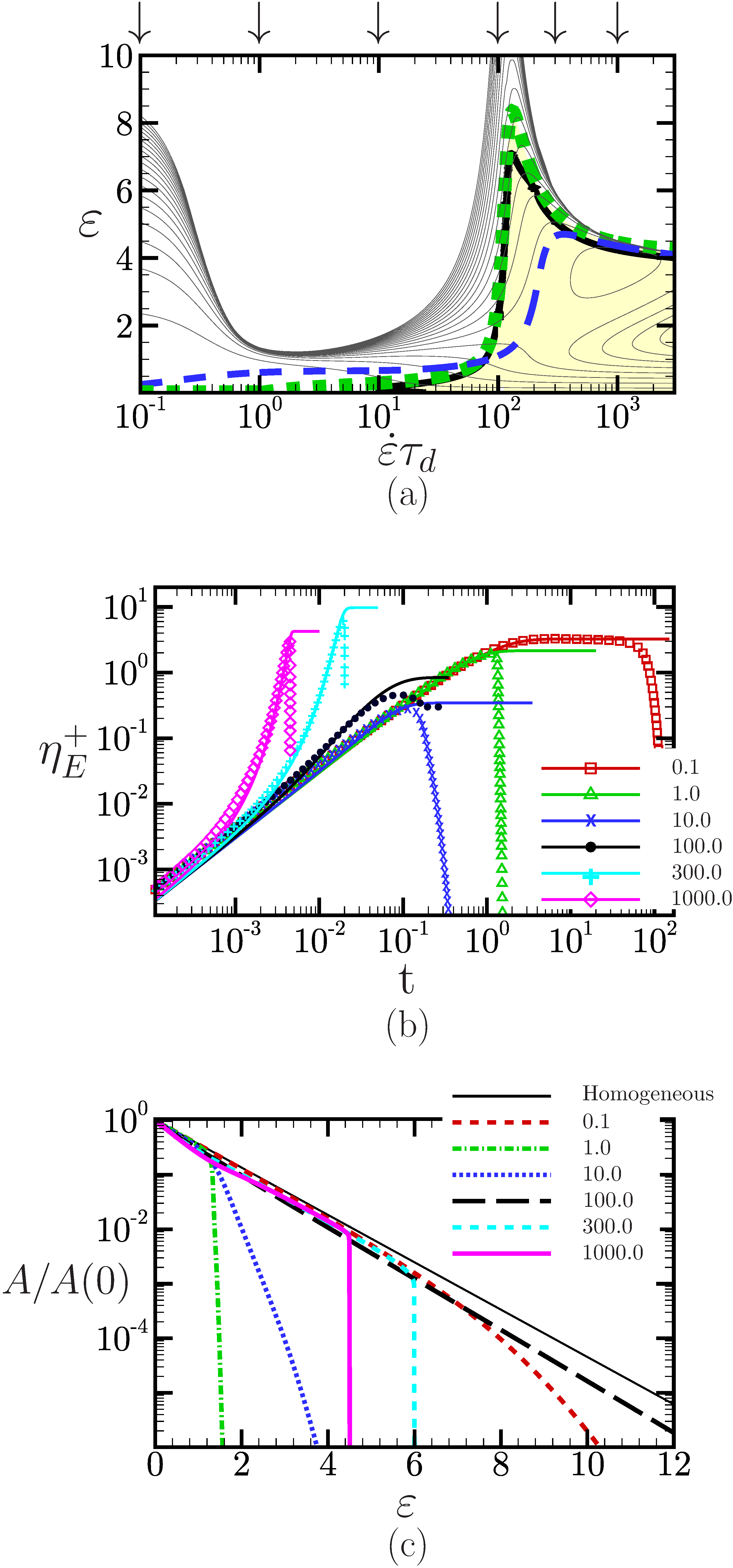}
	\caption{Necking dynamics in nonlinear slender filament simulations of
		the Rolie-Poly model of linear entangled polymeric fluids. (a) Thin
		black lines show contours of the degree of necking heterogeneity.
		Also shown are the criteria of Sec.~\ref{sec:LSA}, calculated using
		values of rheological functions at the filament's midpoint: onset of
		positive eigenvalue to necking (green dotted line), stress
		curvature criterion (thick black line) and original \considere
		criterion (blue long dashed line).  For the six imposed strain rates
		indicated by arrows in (a), the apparent stress growth coefficient
		is reported in (b) for both the nonlinear simulation (symbols) and
		for a calculation in which the filament is artificially assumed to
		remain uniform.  Counterpart results for the filament's cross
		sectional area are shown in (c): assuming homogeneous flow (solid
		black line) and allowing for necking (broken lines).
		\label{fig:NL_RP}}
\end{figure*}

The results of our nonlinear slender filament simulations are shown in
Fig.~\ref{fig:NL_RP}. Panel a) shows the equivalent, for these
nonlinear calculations, of the linear stability data discussed
previously in Fig.~\ref{fig:LSA}f). As before, a vertical cut up this
plane of strain $\epsilon$ and strain rate $\edot$ represents a single
filament stretching run performed at a given imposed strain rate
$\edot$, starting with an unstretched filament at $\epsilon=0$, and
with the filament being progressively stretched out as the accumulated
strain $\epsilon$ increases upward.  Recall the accumulated strain
$\epsilon$ is trivially related to the time $t$ since the inception of
the flow by the relation $\epsilon=\edot t$, in this constant
strain rate protocol. In what follows, we sometimes find it
convenient to refer to time and sometimes strain.

The thin black lines in Fig.~\ref{fig:NL_RP}a) show contours of
constant $\Lambda(t)\equiv A_{\rm hom}(t)/A_{\rm mid}(t)$, where
$A_{\rm hom}(t)$ is the filament area calculated at any time $t$ by
supposing the filament were stretching in a purely uniform way, and
$A_{\rm mid}(t)$ is the actual cross sectional area at the filament's
midpoint.  In this way, $\Lambda=1$ corresponds to a uniform filament,
and $\Lambda$ progressively increases as the filament necks.  In the
figure the first contour has $\Lambda=1$, and each successive contour
crossing as $\epsilon$ increases upwards at fixed $\edot$ corresponds
to an increase in $\Lambda$ by a factor $4^{1/20}$, such that the
$20th$ contour, which is the final shown, represents $\Lambda=4$.
(Although this ratio of areas is relatively modest we note that the
sample is close to necking by this time, because the global area has
become very small.) The more densely clustered the contour lines
vertically at any fixed $\edot$, therefore, the faster necking occurs
in a filament stretching experiment at that strain rate. We have shown
only the first $20$ contour lines, assuming that the sample will have
failed altogether by the time this contour is attained. (We recall,
however, that our slender filament calculation is not capable of
capturing the details of final pinchoff.)

Also included in Fig.~\ref{fig:NL_RP}a), as in Fig.~\ref{fig:LSA}f),
is a green dotted line showing the strain $\epsilon$ at which the
largest eigenvalue (calculated using the values of the material
functions at the filament's midpoint) first becomes positive in any
filament stretching experiment at a given strain rate $\edot$,
signifying the onset of instability to necking.  The thick black solid
line depicts the strain at which the stress curvature criterion
(\ref{eqn:curvature}) predicts this onset. For strains below this
black line, the true tensile stress (calculated at the filament's
midpoint) is an upwardly curving function of time (or accumulated
imposed strain). For strains above it, the stress curves downward in
time.  As can be seen, our stress curvature criterion performs well in
predicting the onset of necking (rapid crossing of $\Lambda$
contours).  Some contour crossing (growth in $\Lambda$) does however
occur before true onset, mainly because of the endplate effects
discussed above.

Comparing Fig.~\ref{fig:NL_RP}a) with Fig.~\ref{fig:LSA}f), we see
that our simplified linear calculation (which assumed a perfectly
uniform base state prior to the onset of necking, together with
simplified periodic boundary conditions), already performed rather
well in predicting the onset of necking in the full nonlinear
calculations. All the features of Fig.~\ref{fig:LSA}f) are preserved
in Fig.~\ref{fig:NL_RP}a), with any small quantitative differences
being explained by the fact that the material functions at the
filament's midpoint in the nonlinear simulations differ slightly from
those in an initially perfectly uniform filament.

The blue long-dashed line in Fig.~\ref{fig:NL_RP}a) shows the strain
$\epsilon$ at which the tensile force overshoot occurs. For strains
below this line, the force increases with strain.  For strains above
it, the force decreases and the original \considere criterion predicts
instability to necking.  As can be seen, the \considere criterion
performs poorly compared to our stress curvature criterion. For
example, it fails to predict the pronounced nose-shaped window of
prolonged stability at imposed strain rates $\epsilon=O(1/\taur)$.  It
also contains no information about the rate of growth beyond onset,
which our calculation provides.

\begin{figure*}
	\centering
	\includegraphics[width=0.75\textwidth ]{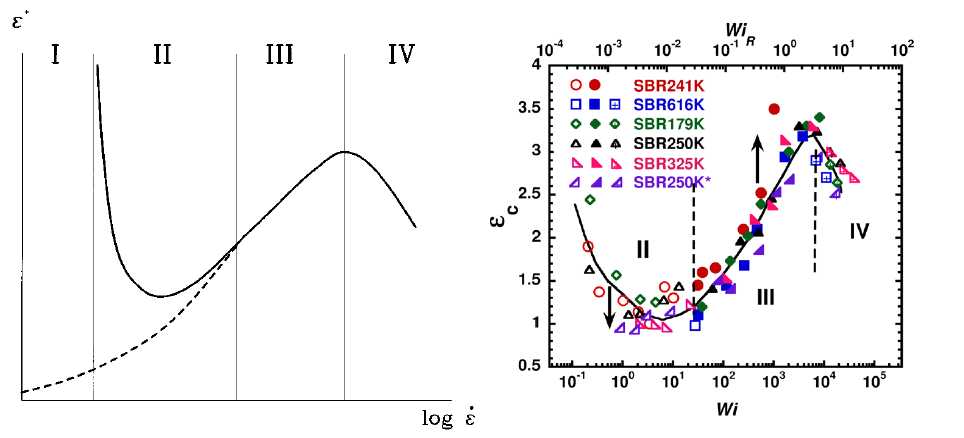}
	\caption{Left) Sketch of strain to failure in filament stretching as a function of imposed strain rate, from \cite{Malkin1997a}. 
		Right) Master curve of experimental data showing strain to failure
		in filament stretching as a function of imposed strain rate, from
		\cite{Zhu2013}.  (Strain rate shown in units of inverse reptation
		time, as in our calculations.
		\label{fig:MalkinPetrie1997}}
\end{figure*}

The necking dynamics at the six imposed strain rate values denoted by
over-arrows in Fig.~\ref{fig:NL_RP}a) are studied in detail in panels
b) and c). At each strain rate, the symbols in panel b) show the
time-evolution of the apparent extensional stress growth coefficient
$\eta_{\rm E,app}^+(t) = F(t)/\edot A_{\rm hom}(t)$. Recall that this
is defined as the tensile force $F(t)$ normalised by the constant
imposed strain rate $\edot$ and the time-evolving cross sectional area
$A_{\rm hom}(t)=A(0) L(0)/L(t)=A(0)\exp(-\epsilon)$ as calculated
supposing uniform stretching, without necking or endplate effects.  It
is this measure of extensional stress that would be reported in a
filament stretching experiment that did not explicitly measure changes
in the filament's cross sectional area due to necking.  For comparison
we also show by solid lines in panel b) the stress growth coefficient
calculated by assuming (incorrectly) that the filament remains uniform
during stretching. As can be seen, when necking sets in the measured
apparent stress growth coefficient decreases compared to that of the
purely homogeneous calculation, because of a reduction in the force
$F(t)$ as the filament thins in the neck.  Such curves show a close
resemblance to experimental ones, for example in 
\cite{Barroso2010}.  For the same six values of imposed strain rate,
panel Fig.~\ref{fig:NL_RP}c) shows the time-evolution of the cross
sectional area at the filament's midpoint as a function of accumulated
strain.  Also shown by a solid line for comparison is the area
evolution $A(t)=A(0)\exp(-\epsilon)$ supposing perfectly uniform
stretching.

Recalling our discussion in Sec.~\ref{sec:results_lsa} above about the
four different regimes of necking dynamics predicted by the Rolie-Poly
model, the results of Fig.~\ref{fig:NL_RP} can be summarised as
follows.

For small imposed strain rates $\edot < 1/\taud=1$ (in our units) the
stress curvature criterion is satisfied, and filament is unstable to
necking, straight away from the inception of the flow. However the
rate of development of the neck is rather modest, $O(\edot)$, giving
relatively widely spaced contours of constant $\Lambda$ in
Fig.~\ref{fig:NL_RP}a): the cross sectional area $A_{\rm mid}(t)$ at
the filament midpoint (red dashed line in Fig.~\ref{fig:NL_RP}c)
deviates relatively gradually from the area $A_{\rm hom}(t)=A(0)
L(0)/L(t)$ that would be expected assuming purely uniform stretching
(black solid line), and a strain $\epsilon\approx 6$ is attained
before significant necking occurs.  In contrast, the tensile stress
attains a steady state on the homogeneous constitutive curve after a
strain of only $\epsilon\approx 1$ at these slow imposed strain rates.
A meaningful measurement of the constitutive curve can therefore be
taken before necking occurs: the red symbols in Fig.~\ref{fig:NL_RP}b)
display a window of steady state between $t=10$ and $30$ before the
stress then falls due to necking.

For imposed strain rates in the regime $1/\taud < \edot < 1/\taur$ the
stress curvature criterion is again satisfied, and the filament is
unstable to necking, right from the inception of the flow. However in
this regime the neck develops much more rapidly, as seen by the
closely spaced contours in panel a), consistent with the sudden fall
at a strain $\epsilon\approx 1.5$ in Fig.~\ref{fig:NL_RP}b) of the
actual cross sectional area at the filament's midpoint (green line)
from the area evolution calculated supposing uniform stretching (black
line).  As discussed in Sec.~\ref{fig:LSA}, this violent necking
behaviour is consistent with the flat region in the material's
underlying constitutive curve in this regime of strain rates
(Fig.~\ref{fig:constitutive}f), associated with the saturation of
chain orientation for $\edot>1/\taud$ and the absence of any
significant chain stretch for $\edot < 1/\taur$.  These two factors
combine to place strong limitations on any tensile stress that the
polymer can provide to restabilise any developing neck. As a result of
this rapid necking, the tensile stress does not have time to attain a
steady state on the homogeneous constitutive curve before significant
heterogeneity develops: the green symbols depart from the green line
in Fig.~\ref{fig:NL_RP}b) before steady state is attained.

At larger strain rates $O(1/\taur)$, significant chain stretch can
develop and provide additional tensile stress, consistent with the
steep slope of the underlying constitutive curve in this window of
strain rates in Fig.~\ref{fig:constitutive}f).  This additional stress
provides some stabilisation against necking and the onset of necking
is deferred until a finite strain has accumulated after the inception
of the flow, as seen by the `upward nose' in the green and black lines
in Fig.~\ref{fig:NL_RP}a). This is consistent with the data in
Fig.~\ref{fig:NL_RP}c) for an imposed strain rate $\edot=100.0$, which
shows the area at the filament's midpoint remaining rather close to
that calculated assuming uniform deformation even up to a strain
$\epsilon=8.0$. (Indeed, some stabilisation effect due to chain
stretch can already be seen at a strain rate $\edot=10.0$, by
comparing the blue and green curves in Fig.~\ref{fig:NL_RP}c.)
However, in this regime of strain rates a typical strain
$\epsilon\approx 10.0$ would be required for the tensile stress to
attain the homogeneous constitutive curve (in a thought experiment
where the filament remained uniform), as seen by the black solid line
in panel b). Because of this, a meaningful measurement of the
constitutive curve still cannot be taken before the filament necks and
the stress declines (black symbols in panel b) compared to that in the
homogeneous calculation. Note that the overshoot shown by the black
symbols in panel b) is in the {\em apparent} stress growth
coefficient, and stems from the decline in tensile force as the
filament fails at the neck.  The Rolie-Poly model lacks any overshoot
in the {\em true} stress growth coefficient (Fig.~\ref{fig:stress}).

Finally for $\edot > 1/\taus$ the chain stretch saturates and the
additional tensile stress that can be provided by this mechanism
levels off, as seen in the constitutive curve in
Fig.~\ref{fig:constitutive}f) at high strain rates. Therefore the
polymer is once again unable to supply additional stress to
restabilise any developing neck, leading to more sudden filament
failure (cyan and magenta symbols in panel b) and lines in panel c).

In the experimental literature, the strain at which a sample fails is
often discussed in terms of a master curve, plotted as a function of
imposed strain rate. This was introduced as a
sketch~\cite{Malkin1997a,Malkin2014}, which we reproduce here in
Fig.~\ref{fig:MalkinPetrie1997}(left). Experimental
data~\cite{Luap2005, Zhu2013} were collected into such a curve for
a styrene-butadiene random (SBR) copolymer linear melt
in~\cite{Zhu2013}, reproduced here in
Fig.~\ref{fig:MalkinPetrie1997}(right). The four different regimes in
this master plot were suggested by by~\cite{Malkin1997a,Malkin2014}
to be interpreted as I ``flow''; II ``transition''; III ``rubbery''
and IV ``glassy''. Based on our results in
Figs.~\ref{fig:constitutive}f), ~\ref{fig:LSA}f) and ~\ref{fig:NL_RP},
however, we now suggest that these four regimes can in fact be fully
interpreted within the Rolie-Poly model as I: a slow flow regime with
gentle necking for $\edot<1/\taud$; II: a regime in which the chain
orientation saturates, leading to fast necking for
$1/\taud<\edot<1/\taur$; III: a regime of increasing molecular
stretch, which provides some stabilisation against necking for
$\edot=O(1/\taur)$; and IV: a regime in which the molecular stretch
saturates, leading again to more rapid necking for $\edot>1/\taur$.
In particular the form of a representative contour in
Fig.~\ref{fig:NL_RP}a), characterising the typical strain to failure,
is in overall agreement with the experimental curves in both subfigure
shown in Fig.~\ref{fig:MalkinPetrie1997}.

\subsection{Branched polymers}
\label{sec:branched}

We now turn to the necking dynamics of branched polymers as modelled
by the Pom-pom constitutive equation of App.~\ref{app:pom-pom}. We
perform calculations separately for the two sets of model parameters
listed in that appendix: PP1, describing a highly branched sample with
a number of arms $q=40$ attached to each end of the molecular
backbone; and PP2, with $q=5$.

\begin{figure*}
	\centering
	\includegraphics[]{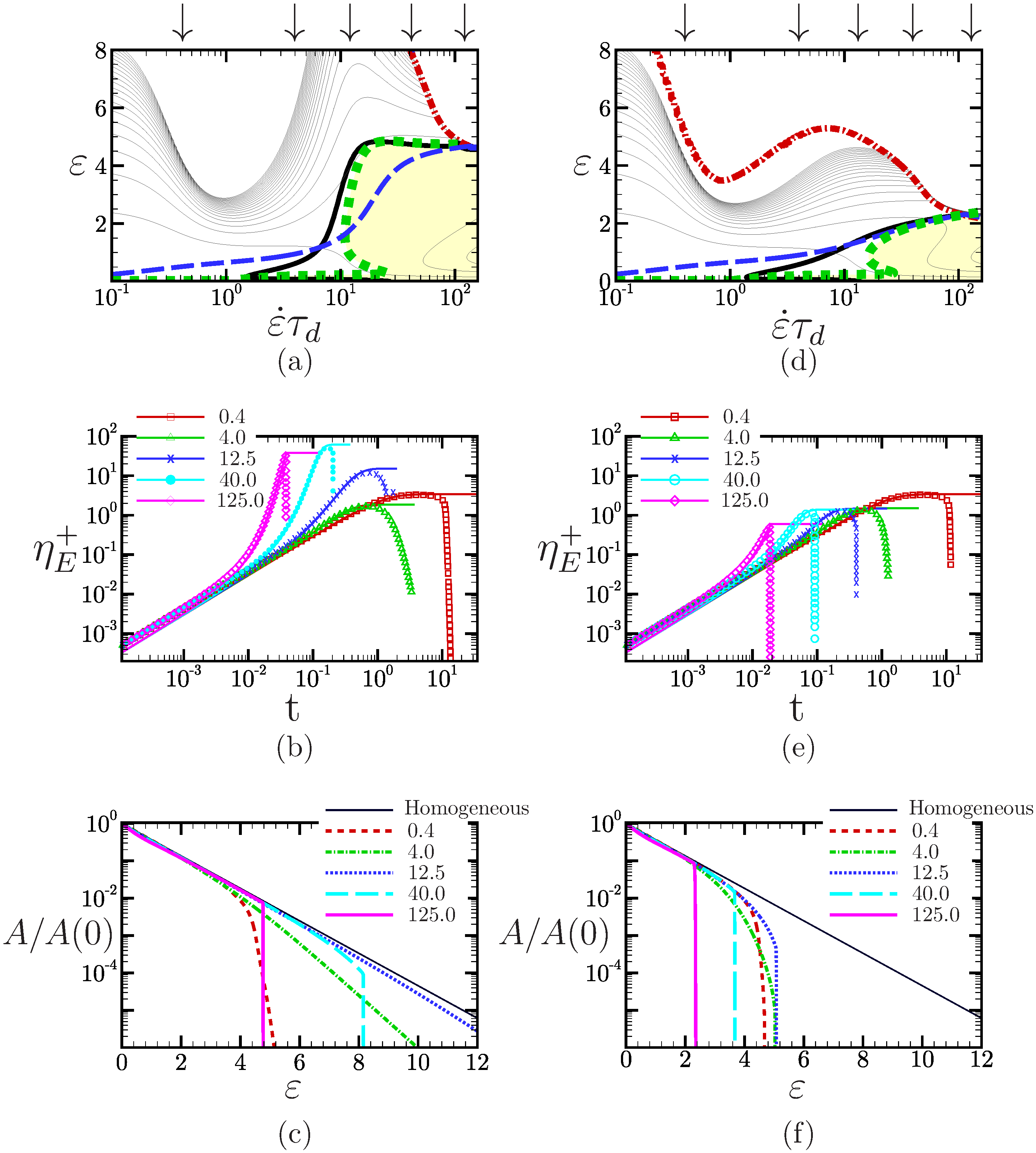}
	\caption{Counterpart of Fig.~\ref{fig:NL_RP}, now for the Pom-pom model of branched polymers. 
	Sub-figure (a)-(c) pertain to sample PP1 and sub-figures (d)-(f) to sample PP2. 
	Model parameter values for each sample are given in App.~\ref{app:models}.
		\label{fig:NL_PP1}}
\end{figure*}

We consider first PP1, for which the underlying homogeneous
extensional constitutive curve was shown in
Fig.~\ref{fig:constitutive}g), and the predictions of our linear
stability calculations were shown in~\ref{fig:LSA}g).  The counterpart
results of our full nonlinear slender filament simulations are shown
in Fig.~\ref{fig:NL_PP1}(a,c,e), in the corresponding format to the
results of Fig.~\ref{fig:NL_RP} for the Rolie-Poly model.

As can be seen, the linear stability calculation of
Fig.~\ref{fig:LSA}g) again performs rather well in predicting the full
nonlinear necking dynamics of the Pom-pom model, despite having
considered the simpler case of small perturbations about a perfectly
uniform filament with periodic boundary conditions. As is evident, the
stress curvature criterion determines the onset of necking for low and
moderate strain rates, before the elastic \considere criterion takes
over at high strain rates.  Furthermore, the four regimes of strain
rate discussed above in the context of linear polymers are mirrored
here for the branched polymers, as follows.

For imposed strain rates $\edot<1/\taub$, where $\taub$ is the
timescale for reorientation of the chain backbone, we see a regime of
relatively gentle necking instability that is effective right from the
inception of flow, triggered by the stress curvature criterion: the
stress signal curves downward as a function of time for all times
$t>0$, as seen by the thick black line at $\epsilon=0$ for these
strain rates.

For the window of imposed strain rates $1/\taub < \edot< 1/\taus$,
where $\taus$ is the timescale for relaxation of backbone stretch, we
find more rapid necking. This is due to the fact that backbone
reorientation has saturated in this regime and is therefore unable to
provide a counterbalancing tensile stress in any developing neck.
This is consistent with the flat region in the underlying constitutive
curve of Fig.~\ref{fig:constitutive}g) in this window of strain rates
$1/\taub < \edot< 1/\taus$. This window is however smaller for PP1 in
Fig.~\ref{fig:NL_PP1}a) than in the Rolie-Poly calculations of
Fig.~\ref{fig:NL_RP}, because we assume that the higher drag afforded
by the dangling arms slows down the relaxation of the backbone stretch
relative to reorientation.  The strain at which significant necking
occurs for the most unstable strain rate is accordingly around $3.0$
in PP1, compared to around $1.5$ for the Rolie-Poly model.

For imposed strain rates $\edot =O(1/\taus)$ significant backbone
stretch develops, consistent with the strong slope in the underlying
constitutive curve of Fig.~\ref{fig:constitutive}g) for such strain
rates. As in linear polymers, this can provide an additional tensile
stress and act temporarily to stabilise the filament against necking
over a finite window of accumulated strain after the inception of
flow. This is evident in the upswing in the black solid and green
dashed lines at $\edot\taub\approx 10.0$ in Fig.~\ref{fig:NL_PP1}a).
At small times (or accumulated strains) the stress curves upward as a
function of time (or strain), consistent with the filament being
stable against necking. Only after a finite strain has accumulated
does the stress curve downward and trigger instability.

For the highest strain rates $\edot>\taus$ the chain stretch
saturates, the underlying constitutive curve flattens out, and the
stabilisation mechanism just described is lost.  Indeed, in the
original version of the Pom-pom model a hard cutoff is imposed in the
degree of backbone stretch that can develop, with the stretch taken to
be entropically limited by the branching priority, $\lambda \le q$.
This pronounced nonlinearity results in the elastic \considere mode of
necking instability setting in at these high strain rates, at a strain
denoted by the red dot-dashed line in Fig.~\ref{fig:NL_PP1}). Once
active, this mode leads to to catastrophically fast necking.

Consistent with this discussion of the different necking regimes, in
the time-signals of the apparent stress growth coefficient and the
cross sectional area at the filament's midpoint
(Figs.~\ref{fig:NL_PP1}b,c) we again see relatively smooth necking at
low strain rates, with much more violent failure at high strain
rate.

As in the Rolie-Poly model, the original \considere criterion (blue
long dashed line in Fig.~\ref{fig:NL_PP1}a) performs less well
compared to our calculations in determining the onset of necking. For
example, it cannot predict the pronounced upswing of stability against
necking around $\edot=10.0$ evidenced by our stress curvature
criterion (black solid line) and eigenvalue calculations (green dashed
line).  It says nothing of the rate at which a neck develops once
instability sets in, and thereby fails (for example) to distinguish
between the relatively gentle necking for imposed flow rates
$\edot<1/\taub$ compared to the faster necking for $1/\taub < \edot <
1/\taus$.  Finally, it fails to distinguish between the two different
modes of instability that we predict: stress curvature at low strain
to moderate strain rates, and elastic \considere at high strain rates.
At high strain rates, however, the onset of the original \considere
mode does coincide with that of our elastic \considere mode,
consistent with the material behaving essentially like an elastic
solid in this regime.

A close comparison of Figs.~\ref{fig:LSA}g) and~\ref{fig:NL_PP1}a)
reveals a greater effect of the elastic \considere mode in the
nonlinear simulations than in the linear ones, with the red dashed
line extending to lower imposed strain rates in
Fig.~\ref{fig:NL_PP1}a). The reason for this is as follows.  In the
linear calculation, the background strain rate used to calculate the
instability criteria is fixed throughout the whole run, with any
necking perturbations that arise being infinitesimal in comparison (by
definition of the linearisation). In contrast, the nonlinear
calculation does account for changes to the rheological quantities in
the developing neck.  In particular, because the neck (by definition)
thins faster than the surrounding material, the fluid within it becomes
subject to a larger strain rate than the globally imposed one.  After
some time of necking via the stress curvature mode, therefore, it
suffers elastic \considere at a lower globally imposed strain rate
than would be predicted by the linear calculation alone.

Results for sample PP2 with fewer arms, $q=5$, are shown in
Fig.~\ref{fig:NL_PP1}(d)-(f). The same four regimes as for PP1 and the
Rolie-Poly model are again evident, though with slightly less well
separated features. The shift in the elastic \considere threshold
compared to that in the linear calculations is particularly pronounced
in this case, due to the much stronger effect of the hard cutoff in
$\lambda$ for this smaller $q$.  In consequence, after an interval of
necking via the stress curvature mode, the region of material in the
neck would eventually succumb to the elastic \considere mode at any
value of the globally imposed strain rate. For strain rate less then
around 20.0, however, this may be unimportant because the sample will
in practice have failed altogether via the stress curvature mode
before the elastic \considere mode has chance to set in.

\begin{figure}
	\centering
	\includegraphics[width=0.5\textwidth ]{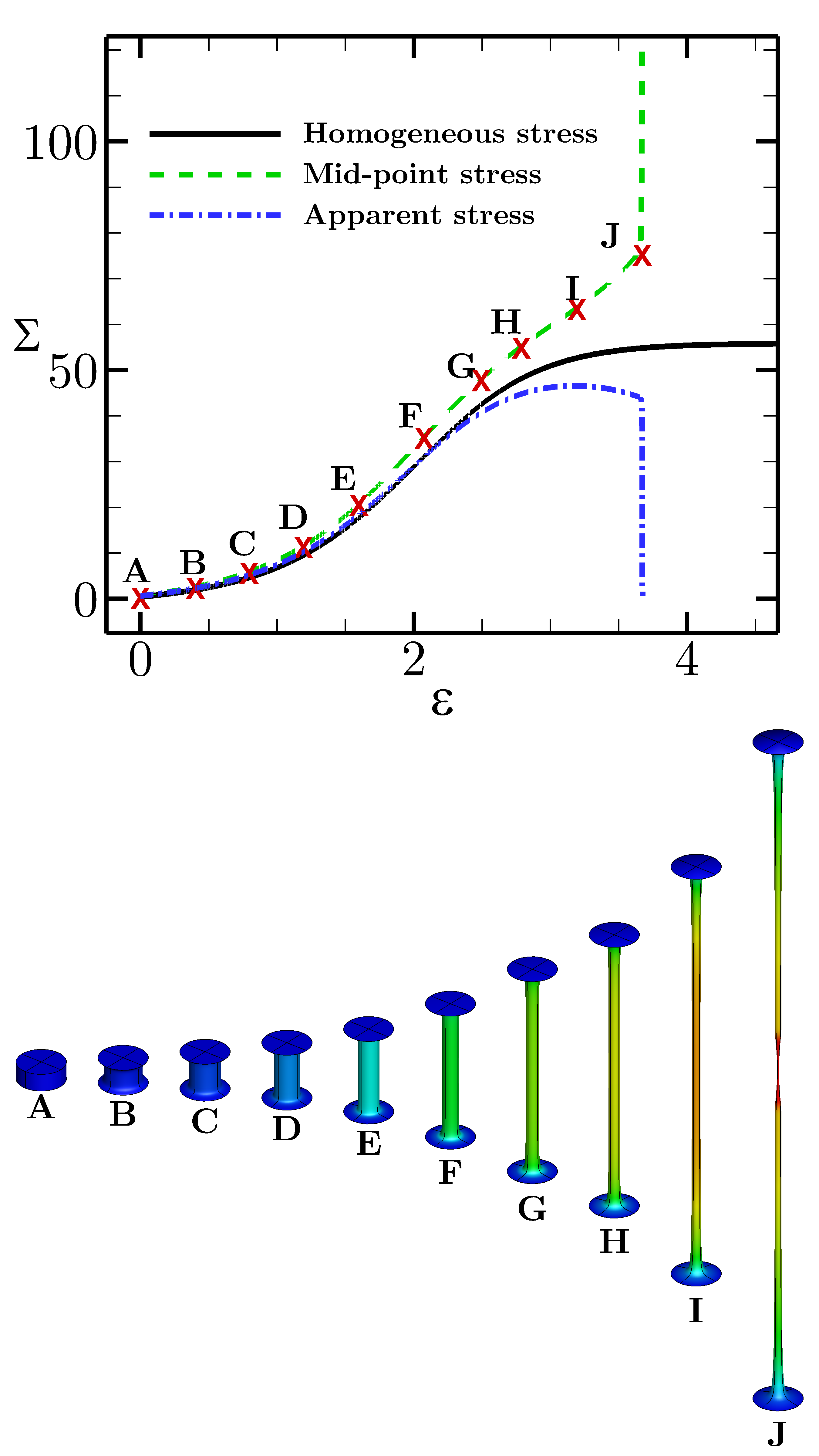}
	\caption{(a) Apparent (dash-dot line) and true mid-point (dotted line) extensional stress growth coefficients versus time in the Pom-pom model (PP2) at a imposed strain rate $\edot=40.0$. Solid line shows the same quantity in a calculation that artificially constrains the filament to remain uniform. (b)  Profile snapshots at the times indicated in (a), with the colourscale showing the tensile stress.
          \label{fig:PP2-profiles}}
\end{figure}

The detailed necking dynamics for $q=5$ at an imposed strain rate
$\edot=40.0$ are shown in Fig.~\ref{fig:PP2-profiles}. Panel a) shows
as dashed, dash-dot and solid lines respectively the mid-point true
stress, apparent stress and (artificial) homogeneous stress growth
coefficient as a function of accumulated strain $\epsilon=\edot t$ in
our necking simulation.  (Recall the definition of these coefficients
at the end of Sec.~\ref{sec:intro}.)  Panel b) shows snapshots of the
filament's profile, with the tensile stress represented by the colour
scale.  (Although the profiles are graphically rendered in 3D, recall
that our calculations are performed in 1D.) As can be seen, at early
times (profiles a-f), the filament area and tensile stress are uniform
along filament, apart from in a small region due to the (mimicked)
no-slip condition near each endplate. Once the stress curvature mode
sets in at a strain $\epsilon\approx 2.0$ the stress profile starts to
become inhomogeneous. Finally, the elastic \considere criterion is
satisfied at the filament's midpoint around a strain $\epsilon\approx
3.2$ and the sample quickly fails.

\begin{figure}
\centering
\includegraphics[width=0.45\textwidth ]{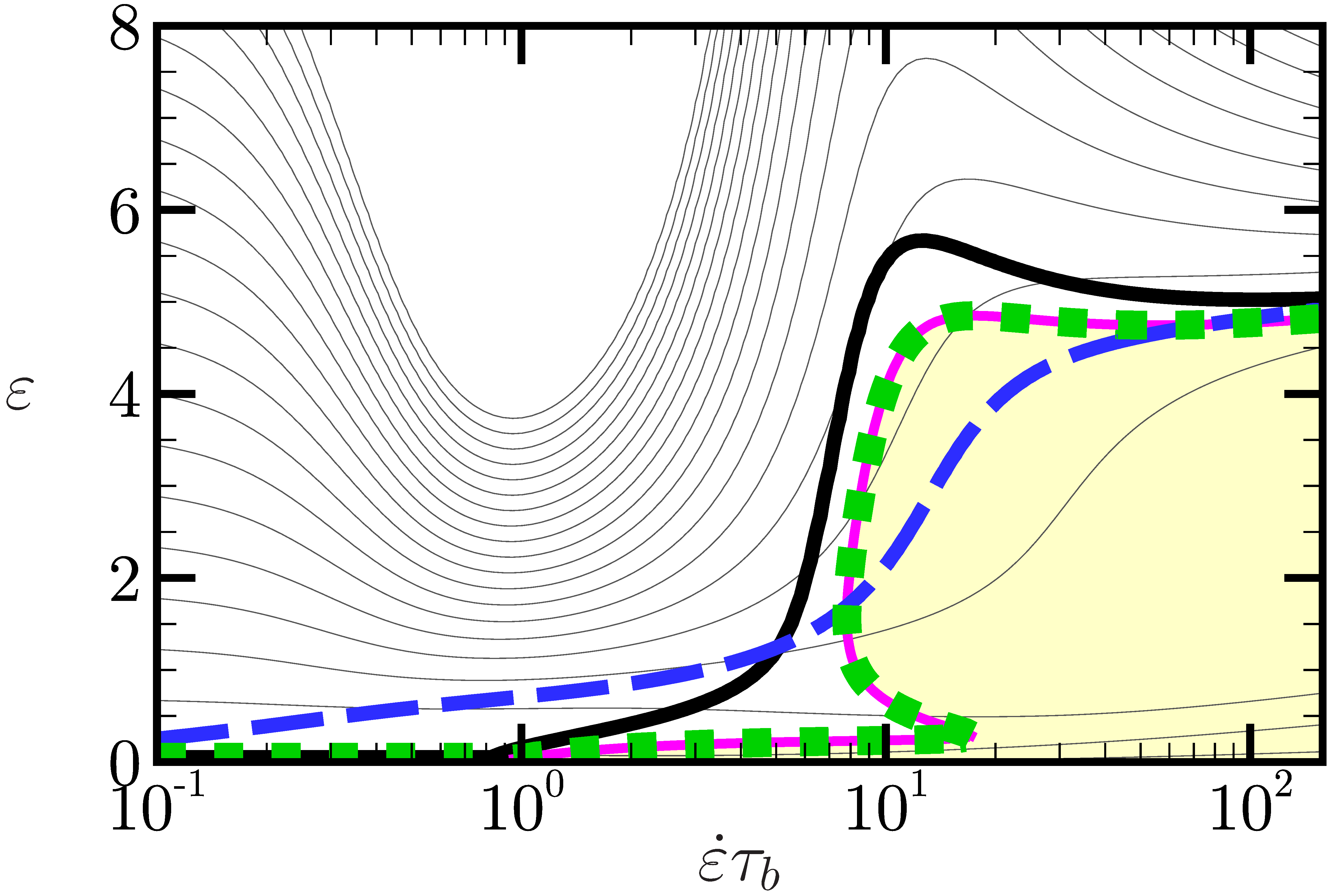}
\caption{Counterpart to Fig.~\ref{fig:NL_PP1}a) for the Pom-pom model
  with parameter values PP1, but with the cutoff condition on chain
  stretch removed.}
  \label{fig:PP_no-finite}
\end{figure}

It is worth noting, however, that the presence of a true elastic
\considere mode in the Pom-pom model relies on the hard cutoff in
backbone stretch imposed in the original version of the model as
proposed by \cite{Blackwell2000,McLeish1998}. Versions of the model
proposed since, for example in Ref.~\cite{Verbeeten2004}, remove this
upper bound in the backbone stretch.  Therefore for comparison with
our results shown in Fig.~\ref{fig:NL_PP1} for the hard cutoff, we
show in Fig.~\ref{fig:PP_no-finite} counterpart (linear stability)
results (for PP1) with the maximum stretch limit removed as
in~\cite{Verbeeten2004}. As can be seen, this removes the elastic
\considere mode and gives very much gentler necking even at high
imposed strain rates.  However experimental evidence (e.g.
\cite{Barroso2010}) suggests that branched materials do show a rather
sudden rupture at sufficiently high strain rates. This may indicate a
harder saturation in backbone stretch than is accounted for
in~\cite{Verbeeten2004}. Indeed, we suggest that the severity of
necking may help to shed some light on the appropriate form of model
dynamics in this regime.

\section{Conclusions}
\label{sec:conclusions}

We have studied the necking dynamics of a filament of viscoelastic
material subject to uniaxial tensile stretching. By means of a linear
stability analysis performed in a constitutive equation written in
highly general form, we have predicted criteria for the onset of
necking that can be expressed simply in terms of characteristic
signatures in the shapes of the experimentally measured rheological
response functions. Given the highly generalised nature of this
calculation, we suggest that the criteria offered here might be
expected to apply universally to all materials. We have provided
evidence for their generality by showing them to apply in numerical
calculations of six popularly used constitutive models: Oldroyd B,
Giesekus, FENE-CR, Rolie-Poly, Pom-pom and a fluidity model of soft
glassy materials.

Two distinct modes of necking instability are predicted. The first
sets in when the tensile stress signal first curves downward as a
function of the time (or accumulated strain) since the inception of
the flow.  Once active, it causes necking with a relatively gentle
rate of development. The second mode sets in when a carefully defined
`elastic derivative' of the tensile force first slopes downward as a
function of time (or strain), and gives much more violent failure.
Whether this `elastic derivative' has any easily measurable
experimental counterpart remains an open question.  Nonetheless, in
the limit of large strain rate $\edot\tau\to\infty$ the elastic
derivative tends towards the ordinary time (or strain) derivative of
the tensile force, and our elastic \considere criterion reduces to the
widely discussed \considere criterion for necking in solids. An
important contribution of this work, however, is to show that the
original \considere criterion fails to correctly predict the onset of
necking in the regime of viscoelastic flows at finite imposed flow
rates $\edot\tau$. Those parts of the rheology literature that have
discussed necking onset in terms of the \considere criterion might
therefore warrant some reinterpretation.

We have studied in detail the way in which our two modes of necking
instability manifest themselves within the microscopically motivated
Rolie-Poly model of entangled linear polymers and wormlike micelles,
and the Pom-pom model of entangled branched polymers. In particular,
we have demonstrated four distinct regimes of necking dynamics,
depending on the value of the imposed strain rate relative to the
inverse reptation and stretch relaxation times. Our theoretically
predicted curve of strain-to-failure as a function of imposed strain
rate is consistent with master curves reported in the experimental
literature~\cite{Malkin1997a,Malkin2014,Luap2005, Zhu2013}.

Throughout we have made the simplifying assumption that spatial
variations develop only along the length of the filament $z$, within a
one-dimensional (1D) slender filament approach that averages (at any
$z$) across the filament's radius $r$ and assumes no variations in the
angular coordinate $\theta$. A comparison of 1D ($z$) and 2D ($rz$)
simulations of viscoelastic extensional flows of dilute and
semi-dilute polymer solutions was carried out in~\cite{Vadillo2012}
where, surprisingly, the 1D approach was found to compare better with
experiment.  Simulation studies of $\theta$ dependent effects near the
endplates were performed in Ref.~\cite{Rasmussen2001,Bonito2006}.
These are beyond the scope of this work, and we hope would not affect
the necking dynamics well away from the endplates, which has been our
focus here.

This manuscript has concerned necking at constant imposed Hencky
strain rate. In a separate manuscript~\cite{stress} we consider the
protocols of constant imposed tensile stress, and constant imposed
tensile force.

We hope that our
calculations will stimulate experimental work to confirm (or disprove)
the criteria offered here, potentially enabling the rheology community
to move beyond the interpretation of necking in complex fluids in
terms of the \considere criterion for necking in solids. It would be
particularly interesting to determine whether the appearance of an
inflexion point (and subsequent downward curvature) in the tensile
stress signal indeed acts as a trigger for the onset of necking.

{\it Acknowledgements} -- The research leading to these results has
received funding from the European Research Council under the European
Union's Seventh Framework Programme (FP7/2007-2013) / ERC grant
agreement number 279365. The authors thank Gareth McKinley for suggesting 
that they work on this problem; and Ole Hassager, Joe Leeman, and Tom McLeish 
for interesting discussions.

\appendix

\newpage
\section{Constitutive models}
\label{app:models}

\begin{table*}
	\centering
	\begin{tabular*}{0.7\textwidth}{@{\extracolsep{\fill}} c | c  }
		\hline
		\hline
		Model & Non-linear parameters \\
		\hline
		Oldroyd-B & - \\
		Giesekus & $\alpha = 0.001$  \\
		FENE-CR & $\delta = 0.001$ \\
		non-stretch Rolie-Poly & $\beta=0.0$ \\
		stretch Rolie-Poly & $\beta=0.0$, $\delta=-0.5$, $\tau_R = 0.00833$ ($Z=40$) \\
		finite-stretch Rolie-Poly & $\beta=0.0$, $\delta=-0.5$, $\tau_R = 0.00833$ and $f = 0.000625$ ($Z=40$)\\
		Pom-pom (PP1) & $\tau_s = 0.1$ and $q = 40$ \\
		Pom-pom (PP2) & $\tau_s = 0.1$ and $q = 5$ \\
		Doi-Ohta fluidity & $\tau_w = 1000$ and $\mu = 0.1$\\
		\hline
		\hline
	\end{tabular*}
	\caption{Parameter values used in our numerical studies of the constitutive models.  The solvent viscosity is taken as $\eta=0.001$ in all cases. \label{tab::model_params}}
\end{table*}

We now detail the constitutive models used in our numerical
calculations.  As discussed in Sec.~\ref{sec:constitutive} of the main
text, in each case the viscoelastic stress $\visc$ can be written as
the product of a constant modulus $G$ and a tensorial function
$\Sfunc$ of a dimensionless microstructural conformation tensor
$\conf$, together with any other microscopic variables
$\lambda,Q,\cdots$ relevant to the fluid under consideration:
\be
\visc=G\Sfunc(\conf,\lambda,Q,\cdots).
\ee
The dynamics of the conformation tensor is then specified by a
differential equation of the general form
\be
\label{eqn:veceb}
\frac{D\conf}{Dt} = \evolve(\nabla
\vecv{v},\conf,\lambda,Q\cdots),
\ee
with counterpart scalar equations for the dynamics of
$\lambda,Q,\cdots$, of the same differential form.  

For notational convenience we define the Lagrangian derivative
\be
\frac{D\conf}{Dt} = \frac{\partial \conf}{\partial t} + \mathbf{v}\cdot\nabla\conf,
\ee
and the upper convected derivative
\be
\overset{\nabla}{\mathbf{W}} = \frac{D\conf}{Dt} - \conf\cdot\mathbf{K} - \mathbf{K}^T\cdot\conf,
\ee
with velocity gradient tensor
$\mathbf{K}_{\alpha\beta}=\partial_\alpha v_\beta$.

\subsection{Oldroyd B model}
\label{app:OldB}

The phenomenological Oldroyd B model represents each polymer chain in
a dilute polymer solution as a simplified dumbbell comprising two
beads connected by a Hookean spring. The relevant conformation tensor
$\conf=\langle \vecv{R}\vecv{R}\rangle$ is then the ensemble average
$\langle \rangle$ of the outer dyad of the dumbbell end-to-end vector
$\vecv{R}$, which is taken to have unit length in the absence of flow.

The viscoelastic stress
\be
\visc = G\left( \conf - \mathbf{I} \right),
\ee
with a constant modulus $G$ (set by the thermal energy $k_{\rm B}T$,
and the volume density of dumbbells.)  In addition to the Hookean
spring force, each bead also experiences viscous drag against the
solvent \cite{Larson1988}, and stochastic thermal fluctuations.
With these dynamics, the conformation tensor obeys
\be
\overset{\nabla}{\mathbf{W}} = -\frac{1}{\tau} \left( \mathbf{W} - \mathbf{I} \right),
\label{eqn:Maxwell}
\ee
with a characteristic relaxation time $\tau$ (set by the thermal
energy, the bead radius, and the solvent viscosity).

As discussed in the main text, for a sustained imposed extensional
strain rate $\edot> 1/2\tau$ the Oldroyd B model displays an
extensional catastrophe in which the dumbbells stretch out
indefinitely and the extensional stress grows indefinitely. The
constitutive curve is accordingly undefined for $\edot>1/2\tau$.

\subsection{FENE-CR model}
\label{app:fene}

The phenomenological FENE-CR model regularises the extensional
catastrophe of the Oldroyd B model by insisting that the extension of
the polymer chains (dumbbells) must remain finite in practice at all
deformation rates.  It does so by replacing the Hookean spring law
with a non-linear spring law \cite{Chilcott1988}, in which the extensional
stress
\be
\label{eqn:fene}
\visc = G f(\conf) \left(  \conf - \mathbf{I} \right),
\ee
and the conformation tensor obeys
\be
\overset{\nabla}{\conf} = - \frac{1}{\tau} f(\conf) \left( \conf - \mathbf{I} \right).
\ee
Here
\be
f\left( \conf \right) = \frac{L^2}{L^2 - R^2} = \frac{1}{1 - \delta T/3},
\ee
in which $3R^2 = T = tr(\conf)$ and $L$ is a parameter that sets the
maximum length of the polymer chains. We write $L= \delta^{-1/2}$ and
work instead with the parameter $\delta$. (The limit $\delta\to 0$
corresponds to Oldroyd B dynamics with infinite extensibility.)  The
FENE-CR model has a well defined extensional constitutive curve at all
strain rates $\edot$, with a finite limiting extensional viscosity
$G\tau/\delta$ as $\edot\to\infty$

\subsection{Giesekus model}
\label{app:Giesekus}

The phenomenological Giesekus model was developed in an attempt to
model more concentrated polymers solutions and melts, starting from the
simple dumbbell model for dilute solutions. It  invokes an
anisotropic drag such that the relaxation time of a molecule
(dumbbell) is altered when the surrounding molecules (dumbbells) are
oriented~\cite{Larson1988}.  The viscoelastic stress
\be
\visc=G(\conf-\mathbf{I}),
\ee
with the conformation tensor obeying the dynamics
\be 
\overset{\nabla}{\conf} = - \frac{1}{\tau} \left( \conf
  - \mathbf{I} \right) - \frac{\alpha}{\tau} \left( \conf -
  \mathbf{I} \right)^2.
\ee
The parameter $\alpha$ controls the degree of drag anisotropy, and
must lie in the range $0 \le \alpha\le 1$. The limit $\alpha\to 0$
recovers Oldroyd B dynamics. For $\alpha > 0$ the Giesekus model
regularises the extensional catastrophe of the Oldroyd B model and has
a well defined constitutive curve at all extension rates, with a
finite limiting extensional viscosity $G\tau/\alpha$ in the limit
$\edot\to \infty$.  For matched (small) values of $\alpha$ and
$\delta$ the steady state extensional constitutive curves of the
Giesekus and FENE-CR models coincide. As shown in the main text, their
extensional {\em dynamics} also closely correspond, though with some
quantitative differences.

\subsection{Rolie-Poly model of linear polymers} 
\label{app:RP}

The three models presented so far are phenomenological in nature,
based on simple dumbbell descriptions.  The Rolie-Poly
model~\cite{Likhtman2003} provides a more microscopically motivated
description of concentrated solutions and melts of linear polymers. As
discussed in the main text, it is based on the tube theory of Doi and
Edwards \cite{Doi1986} in which any given polymer chain is
dynamically restricted by a confining tube of topological
entanglements with the surrounding chains, and is assumed to refresh
its configuration by the basic dynamical process of reptation, chain
stretch relaxation and convective constraint
release
\cite{Marrucci1996,Ianniruberto2014,Ianniruberto2014a}
(CCR). Incorporating these process into a differential constitutive
equation for the dynamics of $\conf=\langle \vecv{R}\vecv{R}\rangle$,
with $\vecv{R}$ the end-to-end vector of a polymer chain, gives
\begin{widetext}
\be
\overset{\nabla}{\conf} = - \frac{1}{\tau_d} \left( \conf - \mathbf{I} \right) 
  - \frac{2}{\taus\left( 1 - f T/3 \right)}\left( 1 - \sqrt{\frac{3}{T}}\right)
\left[ \conf  + \beta\left(\frac{T}{3}\right)^{\delta}( \conf - \mathbf{I} ) \right].
\ee
\end{widetext}
Here $\taud$ and $\taus$ are respectively the characteristic
timescales of reptation and of chain-stretch relaxation. These are
assumed to have the ratio
\be
\frac{\taud}{\taus}=3Z,
\label{eqn::3Z}
\ee
where $Z$ is the number of entanglements per chain.  

In our numerics we take $Z=40$, corresponding to a well entangled
system as used for example in Ref.~\cite{Auhl2008a}.  The parameter
$\beta$ sets the degree of CCR. (Note that the $\beta$ used here
  is distinct from the one used in the definition of the toy model in
  the main text, which is also repeated in Eqn.~\ref{eqn:toyload}
  below.) Following \cite{Likhtman2003} we take $\beta=0.0$ and
$\delta=-1/2$.
The factor $(1-f T/3)$ accounts for finite chain extensibility, as in
Eqn.~\ref{eqn:fene} for the FENE-CR model. This gives a bounded
tensile polymer stress of $3G/f$ in the limit $\edot\to\infty$.

For highly entangled chains (large $Z$) the chain-stretch relaxes
quickly quickly on the timescale of reptation, $\taus\ll\taud$. For
imposed flow rates $\edot\ll 1/\taus$ it is then convenient to take
the limit $\taus \to 0$ upfront and work with the non-stretching form
of the Rolie-Poly, in which the conformation dynamics obey
\be
  \overset{\nabla}{\conf} 
= -\frac{1}{\taud}(\conf-\mathbf{I})-\frac{2}{3}\mathbf{K}:\conf\,(\conf+\beta(\conf-\mathbf{I})).
\ee
This also recovers the reptation-reaction model of wormlike
micelles~\cite{Cates1990} for
$\beta=0$.

In this non-stretching limit and for $\beta=0.0$ (as used in our
numerics) the quantity $\sigma = W_{zz} - W_{xx}$, which determines
the tensile stress as $G\sigma$, evolves according to
\begin{equation}
\frac{D\sigma}{Dt} = \dot\varepsilon \left( 3 + \sigma - \frac{2}{3}\sigma^2 \right) - \frac{1}{\tau}\sigma,
\label{eqn:apprp}
\end{equation}
This has the basic form proposed for the simplified toy version of the
Rolie-Poly model at the end of Sec.~\ref{sec:criteria} with $\beta$ in
that toy model set equal to 1 here. (Note that $\beta$ in the toy
model written in the main text is different from the $\beta$ used in
this appendix for the tensorial Rolie-Poly model. Note also that the
prefactor to the linear loading term in Eqn~\ref{eqn:apprp} is half
that in the toy model in the main text. This however has no effect on
the qualitative behaviour.)

\subsection{Pom-pom model of branched polymers}
\label{app:pom-pom}

The models discussed so far pertain to polymeric fluids with molecules
of linear topology. We now turn to entangled branched polymers in
which each molecule is assumed to comprise a linear backbone with an
equal number of arms $q$ attached to each end, as modelled by the
Pom-pom model~\cite{Blackwell2000,McLeish1998}.  The relaxation of the arms is
considered fast compared to that of the backbone, and so is not
ascribed its own dynamics but acts as an additional drag, which slows
the relaxation of the backbone. The two dominant relaxation processes
in the Pom-pom model are therefore taken to be backbone reorientation,
with a characteristic timescale $\taub$; and backbone stretch
relaxation, with timescale $\taus$. These two timescales are assumed
to be in the ratio
\be
\frac{\taub}{\taus} = Z_b\phi_b,
\label{eqn::phibZ}
\ee
where $Z_b$ is the number of entanglements along the backbone and
\be
\phi_b = \frac{Z_b}{Z_b + 2qZ_a},
\ee
with $Z_a$ the number of entanglements along each arm such that
$\phi_b$ is the fraction of material in the backbone compared that in
the molecule as a whole.

The viscoelastic stress 
\be
\mathbf{\Sigma} = 3G\lambda^2 \left( \mathbf{W} -  \frac{1}{3}\mathbf{I} \right),
\ee
in which $\conf$ is a conformation tensor characterising the backbone
orientation and $\lambda$ encodes the backbone stretch.
\citet{McLeish1998} modelled the backbone orientation both using the
full Doi-Edwards tensor, and also in a simpler a differential
approximation, which we adopt here, taking:
\begin{equation}
\mathbf{W} = \frac{\mathbf{A}}{\rm{tr}(\mathbf{A})},
\end{equation}
with the dynamics of $\mathbf{A}$ obeying the Maxwell model,
Eqn.~\ref{eqn:Maxwell}, with relaxation time $\taub$.  The backbone
stretch evolves according to
\be
\label{eqn:lambda}
\frac{D\lambda}{Dt} = \lambda \mathbf{K}:\mathbf{W} - \frac{1}{\tau_s}\left( \lambda - 1 \right) e^{\nu^*\left( \lambda - 1 \right) }\;\;\textrm{for}\;\;\;\lambda\le q,
\ee
where $\nu^* = 2/(q-1)$, subject to an initial condition $\lambda(0) =
1$. A hard cutoff is then imposed once $\lambda=q$, because the extent
of backbone stretch is taken to be entropically bounded by the
number of arms attached to each end of the backbone, placing an upper
bound on the tensile stress of $3Gq^2$.  As shown in the main text,
the imposition of the hard cutoff leads to catastrophically fast
necking via the elastic \considere mode in this original version of
the Pom-pom model.  We shall also consider a modified model
\cite{Verbeeten2004} in which the hard cutoff is removed, such that
$\lambda$ obeys Eqn.~\ref{eqn:lambda} for all values, and find much
more gradual necking dynamics.

In the main text we present results for two different sets of model
parameters. The first, which we call PP1, has a high number of arms,
$q=40$. 
We take $Z_b\phi_b=10.0$, such that
$\taus=\taud/10.0$, giving a less wide separation of the reorientation
and stretch times for PP1 than in our studies of the Rolie-Poly model,
consistent with the expectation that stretch arises more readily in
branched polymers due to the drag of the side arms.  The second set of
model parameters, which we call PP2, assumes fewer arms, $q=5$, but
still with $\tau_s = \tau_d/10.0$.

\begin{figure*}
\centering
\includegraphics[width=0.8\textwidth ]{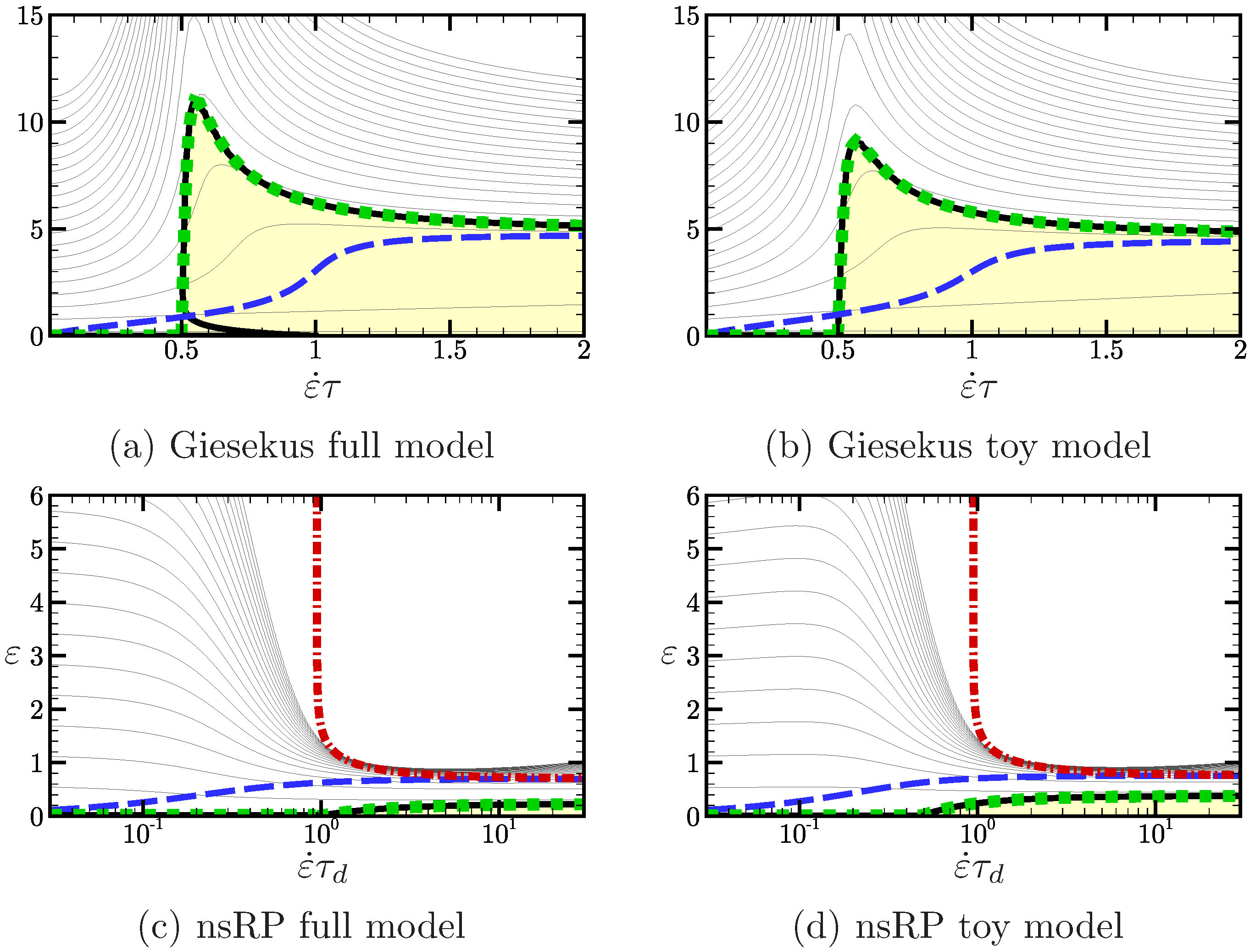}
\caption{Giesekus (top) and non-stretch Rolie-Poly (bottom) full (left) and toy(right)  model comparisons. For both full and toy Giesekus we use $\alpha=0.001$. For the non-stretch Rolie-Poly model this is equivalent to taking $\beta = 2/3$ in the toy model.
  \label{fig:full_toy}}
\end{figure*}

\subsection{Fluidity model of soft glassy materials}
\label{app:fluidity}

We adopt a simplified tensorial fluidity model \cite{Doi1991,
  Hoyle2015} of soft glassy materials~\cite{Sollich1996,
  Cates2003}. This considers a local density function $f(\mathbf{n})$
for the area (per unit volume) of droplet interfaces normal to
$\mathbf{n}$, with a normalisation
\be
Q=\int d\mathbf{n} f(\mathbf{n}).
\ee
The viscoelastic stress 
\be
\Sigma=G\conf
\ee
with a conformation variable
\be
\conf=\int d\mathbf{n} (\mathbf{n}\mathbf{n}-\tfrac{1}{3}\mathbf{I}).
\ee
The conformation variable has the dynamics
\be
\overset{\nabla}{\conf} = \frac{2}{3}Q\mathbf{D} - \conf:\mathbf{K}\left( \frac{2}{3}\mathbf{I} + \frac{\conf}{Q} \right) - \frac{1}{\tau}Q\conf,
\ee
while $Q$ obeys
\be
\frac{DQ}{Dt} = \mathbf{K}:\conf - \frac{1}{\tau}\mu Q^2.
\ee
Here $\mu$ is a phenomenological parameter, which lies in the range $0
\le \mu \le 1$ \cite{Doi1991}.

The relaxation timescale $\tau$ is assigned its own dynamics according
to
\be
\frac{D\tau}{Dt} = 1 - \sqrt{2 \mathbf{D}:\mathbf{D}} \left( \tau - \tau_0 \right),
\ee
where $\mathbf{D}$ is the symmetric part of the velocity gradient
tensor $\mathbf{K}$ and $\tau_0$ is a microscopic time, set to unity
in our units.  The first term on the right hand side captures ageing
in the absence of flow, giving a relaxation timescale $\tau\sim\tw$
that increases as a function of the sample age, {\it i.e.}, the time
since sample preparation (assuming $\tau=\tau_0$ for a freshly
prepared sample). An applied flow (second term) can then arrest ageing
and restore a steady state relaxation timescale set by the inverse
flow rate.

We solve the above equations assuming an initial undeformed sample of
age $\tw$ (such that the initial age $\tau(t=0) = \tw$). 
The initial conditions for $Q$ is a function of $\mu$ \cite{Hoyle2015}.

\subsection{Toy scalar constitutive models}
\label{app:Toy-Model}

At the end of Sec.~\ref{sec:criteria} we introduced a simplified,
generalised, scalarised toy constitutive model that considers only the
tensile component of the viscoelastic stress $\Sigma=G Z$ with
dynamics
\be
\partial_t Z + \vecv{v}.\nabla Z=\edot f(Z)-\frac{1}{\tau}g(Z),
\ee
in which 
\be
f(Z)=3+2Z-\beta Z^2,
\label{eqn:toyload}
\ee
and
\be
g(Z)=Z+\alpha Z^2.
\ee
It was in this model that we analytically derived our criteria for the
onset of necking. We also noted in the main text that this toy model
is capable of giving an excellent approximation to the extensional
necking dynamics of the full tensorial Oldroyd B model
($\alpha=\beta=0$); the Giesekus model ($\alpha\neq 0,\beta=0$); and
the Rolie-Poly model without chain stretch ($\alpha=0,\beta\neq 0$).
This is confirmed in Fig.~\ref{fig:full_toy}, which shows the
counterpart of the results of the full constitutive models of
Fig.~\ref{fig:LSA}, comparing the full tensorial Giesekus model with
its toy scalar counterpart, and likewise the Rolie-Poly model without
chain stretch with its toy counterpart.  This excellent comparison
between the scalar toy models and their full tensorial counterparts
lends additional confidence to our having derived our analytical
criteria for necking within the scalar toy model.

To justify our focusing on just one component of the conformation 
tensor in this toy model, we have checked that in the Giesekus model 
(and finite-stretch Rolie-Poly model) in the vicinity of the pronounced 
nose-shaped curve denoting the onset of instability we have $W_{zz}\gg W_{xx}$, 
justifying our use of just one component $Z=W_{zz}-W_{xx}\approx W_{zz}$ in 
that case.  In the non-stretch Rolie-Poly model the dynamics can exactly be 
written in terms of just one component, $Z=W_{zz}-W_{xx}$, which separately 
justifies our use of just one component in the toy version of that model.  
These facts are consistent with the excellent agreement between our numerical 
results obtained between the full models and their toy counterparts 
in Fig.~\ref{fig:full_toy}.

\end{document}